\documentclass[fleqn,usenatbib]{mnras}

\usepackage{newtxtext,newtxmath}

\usepackage[T1]{fontenc}

\DeclareRobustCommand{\VAN}[3]{#2}
\let\VANthebibliography\thebibliography
\def\thebibliography{\DeclareRobustCommand{\VAN}[3]{##3}\VANthebibliography}

\usepackage{graphicx}	
\usepackage{amsmath}	
\usepackage{float}
\restylefloat{figure}
\usepackage{lipsum}
\usepackage{caption}
\usepackage{placeins}
\usepackage[nottoc]{tocbibind}
\usepackage[usenames]{color}
\usepackage[dvipsnames]{xcolor}
\usepackage{soul}
\DeclareMathAlphabet{\mathpzc}{OT1}{pzc}{m}{it}

\title[Unveiling the origins of galactic bars]{Unveiling the origins of galactic bars: insights from barred and unbarred galaxies}

\author[P. D. López et al.]{
Paula D. López,$^{1}$\thanks{plopez@fcaglp.unlp.edu.ar}
Cecilia Scannapieco,$^{2,3}$
Sofía A. Cora$^{1,4}$
and Ignacio D. Gargiulo$^{4}$
\\
\\
$^{1}$Instituto de Astrof\'isica de La Plata (CCT La Plata, CONICET, UNLP), Paseo del Bosque s/n, La Plata, Argentina\\
$^{2}$Universidad de Buenos Aires, Facultad de Ciencias Exactas y Naturales, Departamento de Física. Buenos Aires, Argentina\\
$^{3}$Consejo Nacional de Investigaciones Científicas y Tecnológicas (CONICET), Buenos Aires, Argentina \\
$^{4}$Facultad de Ciencias Astron\'omicas y Geof\'isicas, Universidad Nacional de La Plata, Paseo del Bosque s/n, La Plata, Argentina
}

\date{Accepted 2024 February 21. Received 2024 February 20; in original form 2023 September 30}
\pubyear{2024}

\begin{document}
\label{firstpage}
\pagerange{\pageref{firstpage}--\pageref{lastpage}}
\maketitle

\begin{abstract}
A significant fraction of local galaxies exhibit stellar bars, non-axisymmetric structures composed of stars, gas, and dust.  Identifying key differences between the properties of barred and unbarred galaxies can uncover clues about the conditions for triggering bar formation.
We explore the early stages of bar formation in a small sample of disc barred galaxies extracted from the TNG50 cosmological simulation, and compare their properties to those of unbarred galaxies.  According to our results, the most important difference between barred and unbarred galaxies is that the former have systematically higher fractions of stellar to dark matter mass in their inner regions, from very early stages and prior to the formation of the bars. 
They harbour high initial gas content, fostering increased star formation rates and leading to a central mass concentration that grows faster over time compared to unbarred galaxies. 
Examining the evolution of the halo spin  within $10\, \mathrm{ckpc}$ reveals that barred galaxies have higher angular momentum transfer from the disc to the halo. Curiously, both barred and unbarred galaxies share similar initial low values of the halo spin, consistent with those proposed in the literature for bar formation.
Furthermore, we evaluate existing stability criteria to capture the complexity of the process, and investigate the effects of mergers, flybys, and environment as possible drivers of bar formation. 
We find no clear link between mergers and disc instabilities resulting in the formation of bars, even though some of the simulated barred galaxies might have been influenced by these events.
\end{abstract}




\section{Introduction} \label{Sec:Intro}

A significant fraction of the local galaxy population exhibits stellar bars. These are central, non-axisymmetric structures made up of stars, gas, and dust following elongated orbits.
Barred galaxies are one of the main families of both lenticular (S0) and spiral (S) galaxies. 
The fraction of barred galaxies depends on whether optical or near-infrared data are used, and whether weak bars are included in the sample. Besides, the dependence of this fraction on galaxy properties as the Hubble morphological type, mass, colour, and bulge prominence is still controversial because of the effect of classification methods in detecting barred galaxies and the different bar definitions \citep[][and references therein]{Lee_2019}. The fraction of barred galaxies in the local Universe amounts to $\sim 47$ per cent when considering all Hubble types later than S0 \citep[e.g.][]{Aguerri_2009,VazquezMata_2022}, although this fraction depends on the specific Hubble type. \cite{Buta_2015} reported estimates of $\sim 55$ per cent over the type range S0/a to Sc, where bars tend to display 3D box/peanut/X patterns, and considerably higher ones ($\sim 81$ per cent) for spiral galaxies within the mid-IR type range of Scd-Sm, where the bars are often linear chains of star-forming regions.

The bar fraction in galaxies also changes over time, influenced by factors such as galaxy mass, luminosity, and colour. A study conducted by \cite{Sheth_2008}  on a sample of luminous face-on spiral galaxies from the 2 deg$^2$ Cosmic Evolution Survey \citep[COSMOS,][]{Scoville_2007} revealed that, in the local Universe, $\sim 65$ per cent of these spiral galaxies contain bars. However, this fraction drops to about $\sim 20$ per cent at $z\sim 0.84$ (a look-back time of 7 Gyr).  
For more massive, luminous, and redder spirals, the bar fraction remains relatively constant up to that redshift.
Conversely, it declines significantly with redshift beyond $z \sim 0.3$ for low-mass, blue spirals.
These trends have received further support from the complementary study conducted by \citet{Melvin_2014}, which extended the analysis to higher redshifts ($0.4 \lesssim z \lesssim 1.0$) considering a sample of massive disc galaxies ($\gtrsim 10^{10}\,M_{\odot}$) with strong bars. Despite variations in the selection criteria between both studies, the overall findings remain consistent.

For quite some time now, studies based on numerical simulations have established that a bar structure can develop in an isolated and self-gravitating cold disc (most of the kinetic energy in rotational motion). The pionering works by \cite{Hohl_1971} and \cite{OstrikerPeebles_1973} followed the evolution of initially balanced rotating discs of stars with an initial velocity dispersion given by Toomre’s local stability criterion against the growth of small-scale irregularities \citep{Toomre_1964}, and found these systems to be promptly and significantly unstable, leading to the emergence of bar-like patterns. The instability initially manifests itself as an open two-arm spiral. As the spiral arms wind up, the elongated structure of a bar emerges in the inner regions of the galactic disc.
Non-axisymmetric perturbations grow through swing amplification resulting from a combination of shear, epicyclic oscillations, and self-gravity \citep{Toomre_1981}. This process starts with a leading spiral moving away from the inner Lindblad resonance (ILR; where the frequency of the perturbation matches the frequency of a star’s epicyclic oscillation).
As it unwinds due to differential rotation, the spiral wave transitions from leading to trailing, with the trailing wave's amplitude becoming much larger than the amplitude of the initial leading wave. The inner part of the perturbation returns to the ILR and is gradually absorbed as it approaches. If this resonance is not present, trailing waves propagate through the galactic center and emerge as leading waves that experience swing amplification, ultimately strengthening the trailing waves. This feedback loop gives rise to a rapidly growing instability (\citealt{BinneyTremaine_2008, Selwood_2014}).

Bar instability can also manifest in disc galaxies through resonant interactions among stars' orbits. These interactions collectively redistribute angular momentum, resulting in the elongation of the galactic structure and the emergence of a bar \citep{LyndenBell_1979}. Additionally, non-spherical structures, such as triaxial bulges and bars, can develop in spherical systems characterised by a prevalence of eccentric orbits (strong radial anisotropy) due to radial orbit instability. The presence of radial orbit instability may be explained by factors such as the inability of velocity dispersion in the transverse direction to resist gravitational forces, linked to the Jeans instability. Another explanation involves the `orbital' approach, extending the concept of Lynden-Bell's bar formation mechanism in disc galaxies to elliptical galaxies, 
which exhibit more radial (round or boxy) orbits, as opposed to the predominantly circular orbits observed in spiral galaxies \citep{Polyachenko_2015, Polyachenko_2020}.

According to the bar formation scenario proposed by \cite{Toomre_1981}, 
disc stability can be achieved by slightly adjusting the inner mass distribution to increase the central angular speed, thereby introducing an ILR that hinders the propagation of density waves, preventing the occurrence of swing amplifications cycles and feedback loops.
\cite{SellwoodEvans_2001} validated this notion through {\em N}-body experiments featuring a quasi-exponential disc and two rigid spherical components, a bulge and a dark matter halo. 
Their research illustrates how galaxies can prevent bar modes by incorporating dense centres; both the disc and bulge contribute significantly to the rotational support in the inner regions, while the (cored) halo has a minor influence. 

These findings contrast with those of \cite{Efstathiou_1982}, who found that bars formed in their simulations with equal strength, regardless of central bulge density, which seemingly contradicts Toomre's prediction.
The conflict between their numerical findings and Toomre's prediction was resolved by \cite{Sellwood_1989}, who demonstrated that highly responsive discs with substantial disturbances can saturate the ILR. This process confines particles within a large-scale bar structure, akin to one that would develop without the presence of a dense center.
Thus, the criterion for stability against bar-like modes in a cold disc presented by \cite{Efstathiou_1982} (ELN criterion) relies simply on the halo-to-disc mass ratio through the expression $\epsilon_{\rm ENL}=V_{\rm max}/(G\,M_{\rm d}/R_{\rm d})^{1/2}$, where $V_{\rm max}$ is the maximum rotational velocity, and $R_{\rm d}$ and $M_{\rm d}$ are the  exponential disc scale length  and total disc mass, respectively.
Their disc models are stable against the growth of bar-like modes  if $\epsilon_{\rm ENL}>1.1$, that is, if they possess a hot component unable to participate in collective instabilities \citep{OstrikerPeebles_1973}, i.e. the criterion is independent of the concentration of the halo or bulge components.
Moreover, this criterion does not consider the significant random motions within the disc or halo, nor does it account for the exchange of angular momentum between the disc and halo, as observed in simulations involving a live halo. Consequently, the ELN criterion may not effectively distinguish between disc galaxies that are stable to bar formation and those that are prone to bar instability
\citep{Athanassoula_2008, Fujii_2018, Devergne_2020, Jang_2023}.

The idea that a high central mass concentration is crucial for preventing bar formation gains further support from subsequent {\em N}-body simulations of disc galaxies, which incorporate live classical bulges and dark matter haloes 
(\citealt{Kataria_2018, SahaElmegreen_2018, Jang_2023}).
These simulations reveal that in galaxies with a massive and compact bulge, bar formation is inhibited. 
The formation of a bar is contingent not only upon a minor concentration of central mass but also on the disc's susceptibility to self-gravitational instability. This susceptibility can be assessed using the Toomre stability parameter, $Q_{\rm T}$ \citep{Toomre_1964, BinneyTremaine_2008}. When this parameter reaches a relatively low minimum value within its radial distribution ($Q_{\rm T,min}\lesssim 1$), swing amplification becomes so pronounced that the inner sections of the spiral arms swiftly coalesce into a bar in less than $1\,{\rm Gyr}$. 
Based on this specific outcome, \cite{Jang_2023} suggest a disc stability criterion that combines data concerning the central mass concentration and $Q_{\rm T,min}$.
In instances where bars do form, the delay in their formation increases with a higher bulge-to-disc mass ratio, and bars tend to be less robust and shorter, and thus
rotate faster, in models featuring more massive and compact bulges, as a result of the development of shorter and stronger ILRs. Besides, angular momentum transfer from a bar to both halo and bulge makes the bar slower and longer over time, consistent with earlier research results \citep[e.g.][]{Athanassoula_1992b, Athanassoula_2003}.
{\em N}-body simulations of disc galaxies, some of them featuring
gas and/or a triaxial halo, reveal that the bar formation and evolution are significantly influenced by the relative gas fraction and the shape, density and spin of the halo
\citep{Athanassoula_2013, Bland-Hawthorn_2023, SahaNaab_2013, Long_2014, Collier_2018, LiShlosman_2023}.

Cosmological hydrodynamic simulations offer a consistent way to investigate the process of bar formation in the context of galaxy growth within a $\Lambda$ Cold Dark Matter cosmology. These simulations have successfully produced barred galaxies with realistic structural, photometric and chemokinematical properties 
(\citealt[][based on Aquarius Project]{ScannapiecoAthanassoula_2012}; \citealt[][using Auriga simulations]{Grand_2017, BlazquezCalero_2020}; \citealt[][considering {\sc eagle} simulations]{Algorry_2017}; \citealt[][using the Eris  simulations]{Spinoso_2017} \citealt[][based on the {\sc CLUES} Project]{Marioni_2022}).
They have also led to a consensus regarding the properties of disc galaxies that give rise to bar formation: these galaxies tend to assemble at earlier times, exhibit larger stellar-to-halo ratios in their inner regions, and possess more massive but smaller and colder discs compared to unbarred galaxies \citep[][Illustris TNG50, TNG100]{Izquierdo-Villalba_2022}.
An analysis of barred galaxies using the TNG100 run reveals that approximately $55$ per cent of disc galaxies with a stellar mass $\sim 10^{10.6}\,M_{\odot}$ host bars, aligning well with observations. However, discrepancies are observed for more and less massive galaxies \citep{Zhao_2020}. Theoretical predictions and observational data find agreement when focusing exclusively on long bars \citep[][Illustris TNG50]{RosasGuevara_2022}. 
The formation and characteristics of bars are significantly influenced by the properties of the host galaxy, including its mass and environment \citep{Lokas_2020, Lokas_2021b}. In turn, the bar itself redistribute stars and gas producing a notable impact on the properties and dynamics of its host galaxy  \citep{Athanassoula_2013Book, Spinoso_2017, Fragkoudi2020}.

The performance of the ELN criterion to determine the stability against bar formation have also been assessed from the analysis of cosmological hydrodynamical simulations.
Recently, from the analisys of TNG100 and TNG50 simulations, \cite{Izquierdo-Villalba_2022} conclude that this criterion effectively characterizes the stability of the majority of massive disc galaxies undergoing secular evolution. Barred galaxies misclassified as stable discs ($\sim 25$ per cent of their sample) may have their bars formed due to external factors, such as encounters with massive satellites or host haloes with high spin parameter.
Recognizing that systems seemingly unstable based on the ENL criterion may find stability within massive haloes characterized by high-velocity dispersion \citep{Athanassoula_2003}, \cite{Algorry_2017} introduced an additional parameter that accounts for the overall significance of the entire system, including its halo. This parameter is defined as the ratio between the circular velocity at the half-mass radius and the maximum circular velocity of the surrounding halo. This combined approach has proven successful in accurately identifying systems that are unstable against bar formation, as further substantiated by the findings of \cite{Marioni_2022}.

From the observational side, utilising a statistically unbiased sample of barred and non-barred galaxies, \cite{Romeo2022} investigate the effectiveness of both the ELN criterion and the mass-weighted Toomre parameter of atomic gas in detecting the presence of bar structures. Their study revealed that the ELN criterion is considerably inaccurate, as it fails in nearly half of the cases analysed. On the other hand, the mass-weighted Toomre parameter succeeds in distinguishing barred from non-barred galaxies only in a statistical sense.

The underlying mechanisms and relative significance of various triggers for bar formation and growth remain enigmatic. Existing criteria for assessing disc stability fail to comprehensively capture the complexities of this process. Most prior studies have taken a statistical approach to this problem. Our present research is dedicated to unveiling the properties and mechanisms involved in the early stages of bar formation, even before a distinct structure emerges. Our main goal is to decipher the fundamental differences that lead to bar formation in disc galaxies. Additionally, we want to evaluate the efficiency of current stability criteria. This integral analysis could serve as a stepping stone toward the development of more accurate criteria. To achieve this, we examine the evolutionary paths of a selected group of barred and unbarred galaxies, extracted from the sample presented by \cite{RosasGuevara_2022}. Our focus is on identifying disparities in properties between barred and unbarred galaxies at various redshifts. Preliminary results have been presented in  \cite{Lopez_2023_baaa}.

This paper is organized as follows. In Section \ref{Sec:Sim&GxSelec}, we briefly summarize the main features of the simulation used in this work and present the properties of the galaxies selected for the analysis, as well as the method used to identify the bars. In Section \ref{Sec:BarProp}, we describe the main properties of barred galaxies at $z=0$: bar strength, length, phase, pattern speed, and corotation radius. In Section \ref{Sec:EvolBarProp}, we define the formation time of bars, present an analysis of the evolution of the bar properties, assess the influence of the dark matter halo spin on bar formation, and explore the connection between bar development and the emergence of a central mass concentration; additionally, we analyse the relative contributions of stars and gas in comparison to the dark matter mass within the central regions.
In Section \ref{Sec5}, we analyse different bar formation criteria based on disc instabilities, and the possible effects of mergers, flybys and environment.
In Section \ref{Sec:Discussion}, we discuss our results and compare them with other theoretical and observational studies. Finally, in Section \ref{sec:conclusions}, we present  the conclusions of our work.

\section{Simulations and galaxy selection} \label{Sec:Sim&GxSelec}

\subsection{ Overview of the TNG Simulations} \label{Sec:TNGSim}

In this work, we make use of simulations from the IllustrisTNG project, a collection of cosmological simulations designed to explore galaxy formation and evolution (\citealt{nelson2018first, naiman2018first, pillepich2018first, marinacci2018first, springel2018first}). This suite comprises simulations with varying volumes, ranging from $50$ to $300$ cMpc (comoving), and features different spatial and mass resolutions. The simulations are based on the cosmological parameters derived from the Planck Collaboration's findings \citep{collPlanck2016}, including , $\Omega_\mathrm{m}=0.3089$, 
$\Omega_{\Lambda}=0.6911$, $\Omega_\mathrm{b}=0.0486$, $\sigma_8=0.8159$, $h=0.6774$ and $n_\mathrm{s}=0.9667$. These parameters describe the average densities of matter, dark energy, and baryonic matter, as well as other essential cosmological values. 

The simulations were conducted using the {\sc arepo} moving-mesh code \citep{Springel_2010}, which accounts for gravitational interactions and incorporates sub-grid models to represent various baryonic processes, building upon earlier work from the Illustris project \citep{Vogelsberger_2014, Genel_2014}. 

The simulation of primary interest in our research is TNG50, notable for its high resolution and specific characteristics (\citealt{TNGNelson2019, Pillepich_2019}). TNG50 features a box size of approximately $50$ cMpc and a dark matter mass resolution of $4.5 \times 10^5 \, \rm{M}_{\odot}$, and a mass resolution for baryonic matter of $8.5 \times 10^4 \rm{M}_{\odot}$. In terms of gravitational softening, stars and dark matter are treated with softening lengths of $288\,{\rm pc}$, while the gas softening is adaptive with a minimum value of $74\,{\rm pc}$ in comoving units.

\subsection{Galaxy sample}\label{Sec:Sample}

This work is mainly focused on the identification of  particular features in the evolution of galaxy properties that are connected with the formation of a bar, which can give us information about the physical processes involved in bar formation. With this aim, we restrict our analysis to two small sets of $z=0$ barred and unbarred disc galaxies, rather than following a statistical approach.

Our selection of barred and unbarred galaxies is based on the $z=0$ classification made for TNG50 simulation by \citet{RosasGuevara_2022}. 
In contrast to the unbarred galaxy population, their barred galaxies exhibit several distinguishing characteristics. They possess an older stellar population, lower gas fractions, and reduced star formation rates, and their discs tend to assemble earlier and at a faster pace. Barred galaxies are commonly found in haloes characterised by higher concentrations and smaller spin parameters. Additionally, the inner regions of barred galaxies are more dominated by baryonic matter, while their overall global stellar mass fractions are similar to unbarred galaxies.
We have randomly chosen eight galaxies from their dataset, with four of them featuring bars. The sole criterion used for this selection is their virial mass. These galaxies fall within the virial mass range
 $3.2 \times 10^{11} {\mathrm{M}}_{\odot}$ to  $1.5 \times 10^{12} {\mathrm{M}}_{\odot}$ at $z=0$. 

\begin{table}
    \centering
    \begin{tabular}{lcccccc}
        \hline
        Galaxy & $M_{\rm{star}}$ & $M_{\rm{gas}}$ & $R_{50}$ & $D/T$ & $M_{\rm{vir}}$ & $R_{\rm{vir}}$ \\
        & $10^{10} \, {\rm M}_{\sun}$ & $10^{10} \, {\rm M}_{\sun}$ & kpc & - & $10^{11} \, {\rm M}_{\sun}$ & kpc \\
        \hline
        B$_1$ & 3.88 & 3.57  & 6.92  & 0.67 & 5.52 & 169.29   \\
        B$_2$ & 4.71 & 2.83  & 7.54  & 0.73 & 5.51 & 169.18   \\
        B$_3$ & 3.67 & 2.61  & 4.0   & 0.64 & 4.16 & 154.06   \\
        B$_4$ & 7.44 & 12.10 & 10.32 & 0.72 & 13.33 & 227.01  \\
        NB$_1$ & 3.51 & 5.89 & 7.66  & 0.85 & 6.90 & 182.30  \\
        NB$_2$ & 6.91 & 8.43 & 10.66 & 0.88 & 10.20 & 207.70\\
        NB$_3$ & 1.07 & 2.14 & 4.45  & 0.85 & 3.28 & 142.29  \\
        NB$_4$ & 2.53 & 5.83 & 8.83  & 0.88 & 4.80 & 161.52  \\
        \hline
    \end{tabular}
    \caption{Global properties of the selected galaxies: the total stellar mass ($M_{\rm{star}}$), the total gas mass ($M_{\rm{gas}}$), the stellar half-mass radius ($R_{50}$) and the disc-to-total mass ratio ($D/T$) of the galaxies; the virial mass ($M_{\rm{vir}}$) and the virial radius ($R_{\rm{vir}}$) of the corresponding subhaloes.}
    \label{tab:gxsProps}
\end{table}

 \begin{figure*}
    \includegraphics[width=2\columnwidth]{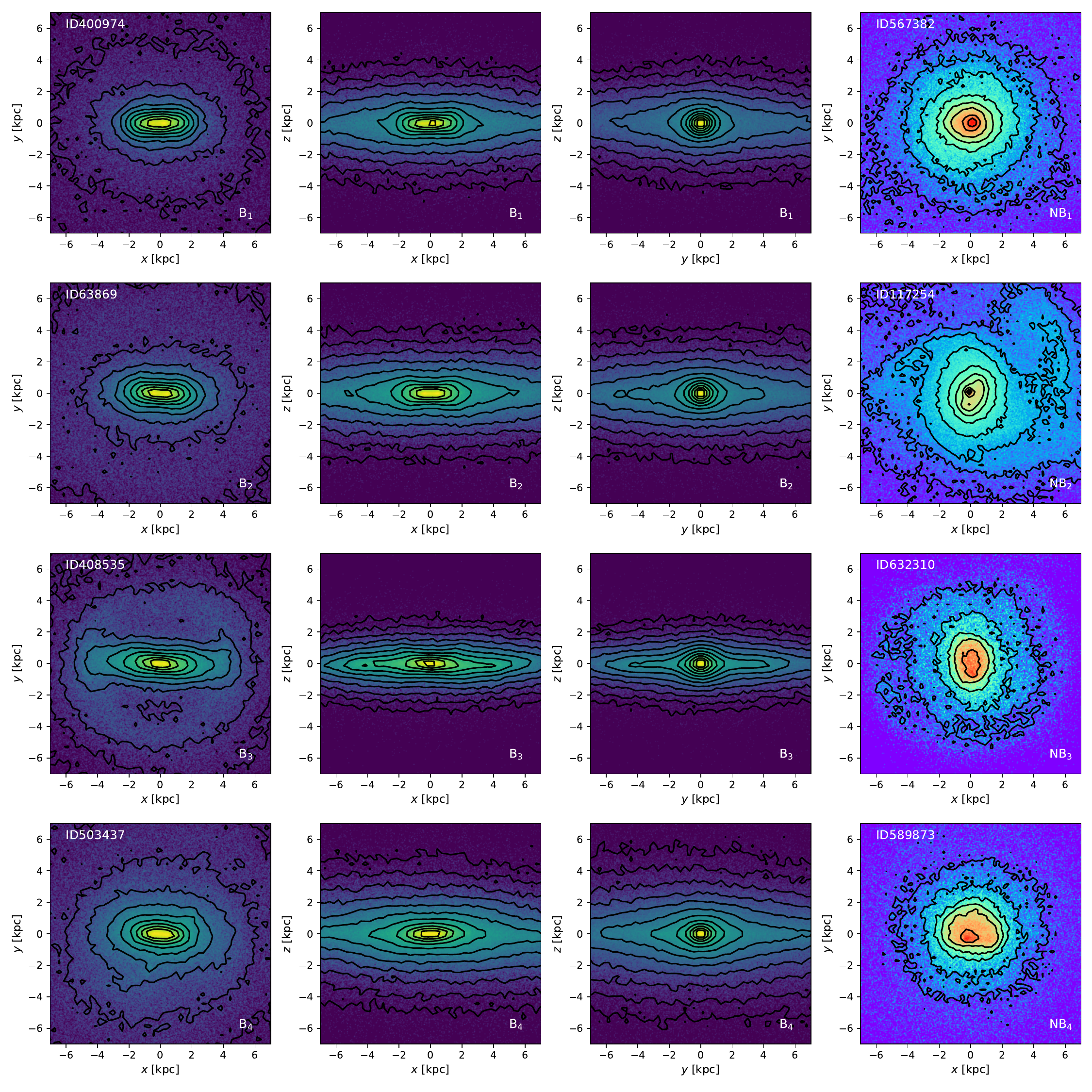}
    \caption{Projections of stellar mass density of the eight galaxies in our sample at $z=0$. The first three columns show three different projections for the four barred galaxies: face-on ($xy$ plane), edge-on ($xz$ plane), and end-on ($yz$ plane). The fourth column shows the face-on projections of the four unbarred galaxies.}
    \label{fig:tot}
\end{figure*}

\begin{figure*}
\includegraphics[width=2\columnwidth]{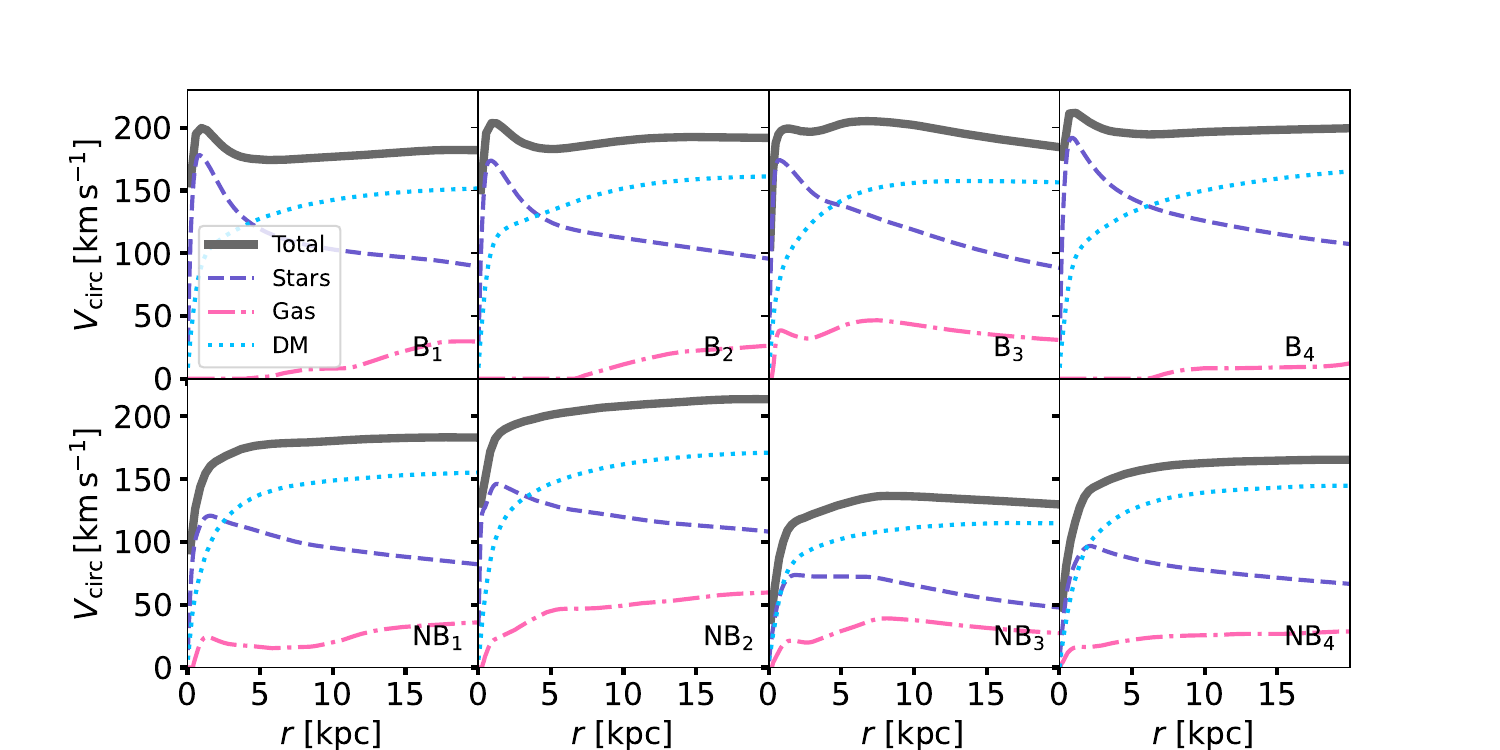}
    \caption{Circular velocity curves (solid lines) for barred galaxies (\textit{upper panels}) and unbarred galaxies (\textit{lower panels}), at $z=0$, as a function of radius $r$. We also show separately the contributions of the dark matter (dotted lines), stellar (dashed lines)  and gaseous (dotted-dashed lines) components. 
    }
    \label{fig:Vcirc}
\end{figure*}

Fig.~\ref{fig:tot} shows maps of the projected stellar mass density, at $z=0$, for the eight selected galaxies: the first three columns correspond to projections with different orientations for the galaxies with bars -- referred to as B$_1$, B$_2$, B$_3$ and B$_4$ -- and the fourth column shows the face-on projections of the unbarred galaxies -- NB$_1$, NB$_2$, NB$_3$ and NB$_4$. The projection of the systems is such that the discs are contained in the $xy$ plane, and the total angular momentum of the stars is aligned with the $z$-axis. All galaxies show a prominent stellar disc, although there is some degree of variation in the disc sizes and particular features of the different galaxies. 
The presence of the bar structure is evident in the three projections of the barred galaxies.

The main characteristics of the selected galaxies, at $z=0$, are detailed in Table~\ref{tab:gxsProps}: total stellar mass ($M_{\rm{star}}$), total gas mass ($M_{\rm{gas}}$), stellar half-mass radius radius ($R_{50}$), disc-to-total mass ratio ($D/T$), virial mass ($M_{\rm{vir}}$) and virial radius ($R_{\rm{vir}}$).
The total stellar and gas masses, $M_{\rm{star}}$ and $M_{\rm{gas}}$, and the stellar half-mass radius radius ($R_{50}$) are given by the simulation catalogue\footnote{https://www.tng-project.org/data/}. The virial radius, $R_{\rm{vir}}$, is defined as the radius where the density equals $200$ times the critical density of the universe, and the virial mass, $M_{\rm{vir}}$, as the total mass enclosed within $R_{\rm{vir}}$.

The $D/T$ ratio is calculated using a kinematic decomposition of the stellar component. As in \cite{Scannapieco_2009}, we define the circularity parameter of each star as $\epsilon = j_\mathrm{z} / j_\mathrm{circ}$, where $j_z$ is the angular momentum perpendicular to the disc plane, and $j_\mathrm{circ} = r \cdot V_\mathrm{circ}$ is the angular momentum corresponding to a circular orbit at the star's radius, with $V_\mathrm{circ}$ the circular velocity of the system. It is worth noting that there are different ways to estimate the $D/T$ ratios (e.g. \citealt{Abadi_2003}) including estimations mimicking observational techniques \citep{Scannapieco_2010}. For our sample, the $D/T$ ratios are similar for the barred and unbarred galaxies, with an average value of $0.69$ for the former and $0.86$ for the latter. Note that although one would expect that barred galaxies have more massive discs,  the bar component would count in the kinematic decomposition as part of the spheroid.
Finally, it is worth noting that both the barred galaxy $\rm{B}_4$ and the unbarred galaxy $\rm{NB}_2$ are significantly more massive than the rest of the galaxies within their respective subsamples and are located within very massive dark matter subhaloes.

Fig.~\ref{fig:Vcirc} shows circular velocity curves as a function of radius,  for barred and unbarred galaxies (upper and lower panels, respectively), at $z=0$. 
This is obtained from the expression $V_{\rm{circ}}(r) = \sqrt{(G \, M(r))/r}$, where $M(r)$ is the total mass inside a radius $r$. 
The maximum circular velocities reached in the innermost $10\, {\rm kpc}$ of the barred galaxies are comprised in the range $\sim 199.7 - 211.8\,{\rm km}\,{\rm s^{-1}}$, consistent with the values found by \cite{Schmidt_2023} from the rotation curves estimated for a sample of 46 barred galaxies in the local Universe of MaNGA (Mapping Nearby Galaxies at Apache Point Observatory) with stellar masses in the range $\sim 2 - 3.4 \times 10^{10}\,{\rm M}_{\odot}$. The maximum circular velocities obtained for our sample of unbarred galaxies span a range of $\sim 136.7 -208.2 \,{\rm km}\,{\rm s^{-1}}$, much wider than for barred galaxies and reaching lower values. 

We also show separately results for the gaseous, stellar and dark matter components ($M(r)$ is estimated from the spatial distribution of the corresponding particles in the simulation), in order to highlight their relative contributions to the total circular velocity.  
We find that, at $z=0$, stars largely dominate the mass content in the inner regions of the four barred galaxies. In contrast, the contribution of stars relative to the dark matter component is lower or subdominant for the unbarred galaxies.  
The gas comprises, in all cases, a small fraction of the total mass and, in the case of three out of the four barred galaxies, there is a negligible amount of gas in the inner $\sim 5~\mathrm{kpc}$.
The situation differs for unbarred galaxies, where the proportion of gas relative to the other components is larger, specially in the innermost regions.

\subsection{Identification of bars} \label{Sec:IdentificationBars}

We identify the bar components in the simulated galaxies by performing a  Fourier analysis of the face-on, stellar projected mass density  $\Sigma(R,\theta)$, as in \cite{ScannapiecoAthanassoula_2012}:
\begin{equation}
    \Sigma(R, \theta)=a_{0}(R)+\sum_{m}\left[a_{m}(R) \cos (m \theta)+b_{m}(R) \sin (m \theta)\right],
\end{equation}
where $R$ is the projected radius (i.e. in the $xy$ plane) and $\theta$ is the azimuthal angle. Each coefficient of the expansion is given by the expressions:
\begin{equation}
    \begin{aligned}
        &a_{m}(R)=\sum_{i} m_{i} \cos \left(m \theta_{i}\right), m \geq 0, \\
        &b_{m}(R)=\sum_{i} m_{i} \sin \left(m \theta_{i}\right), m>0,
    \end{aligned}
\end{equation} 
where $m_i$ is the mass of each star particle and $m$ denotes the mode of the Fourier expansion's terms. The calculation was made by taking rings of $\Delta r = 0.1$ ckpc width,
starting at the center of each galaxy.  The amplitudes of the Fourier modes are defined as  $I_0 = a_0$ and $I_\mathrm{m} = \sqrt{a^2_{\mathrm{m}} + b^2_{\mathrm{m}}}$ for $m>0$.

The strength of the bar can be quantified using the ratio $I_2/I_0$, which we  refer to as the parameter $A_2$, hereafter. In this work, we assume that a barred structure is present if the maximum in the  $A_2$ radial profile, $A_{2}^{\mathrm{max}}$,  exceeds the  threshold $A_{2}^{\mathrm{thresh}}=0.2$,
following a criterion similar to \cite{Algorry_2017}.   
Furthermore, we confirm the existence of the bar at the respective time instances by  visual examination of the surface stellar density projections.

\section{Properties of galaxies and bars at \texorpdfstring{$z=0$}{Lg}}
\label{Sec:BarProp}

\subsection{Strength, phase and length of the bars}
\label{Sec:StrengthPhaseLength}

\begin{figure*}
 \includegraphics[width=2\columnwidth]{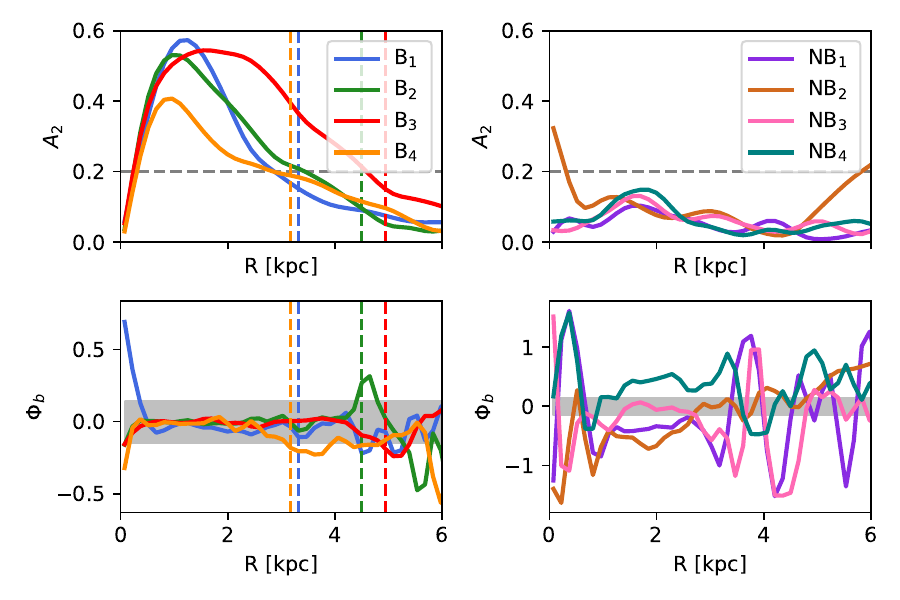}
     \caption{
      {\it Upper panels:} Projected radial profiles of the strength of the 
    bar-like structure
     quantified using the parameter A$_2$
    for galaxies identified as barred (left-hand panel) and unbarred (right-hand panel), at $z=0$. The threshold $A_{2}^{\mathrm{thresh}}=0.2$ of our criterion to identify bar-like structures is shown as an horizontal dashed line. {\it  Lower panels:}  
     Phase of the $m=2$ Fourier mode
     ($\Phi_{\rm{b}}$) as a function of the projected radius, for the barred (left-hand panel) and unbarred (right-hand panel) galaxies, at $z=0$. The region consistent with a constant phase ($|\Delta \Phi_{\rm{b}}| < 0.15$) is highlighted with a rectangular shade. In the case of galaxies with bars, the corresponding bar length is indicated by a vertical dashed line.}    
     \label{fig:a2rad_phirad}
\end{figure*}

The upper panels of  Fig.~\ref{fig:a2rad_phirad} show the strength of the bar, $A_2$, as a function of projected radius for the eight simulated galaxies, at $z=0$, separated into barred (upper left-hand panel) and unbarred (upper right-hand panel) systems. The presence of a bar is clearly evident in the radial profile of $A_2$ for all barred galaxies, with  values well above the threshold $A_{2}^{\mathrm{thres}}$ in the inner regions, up to distances of  $\sim 3 \,{\mathrm{kpc}}$.
In contrast,  the radial profiles of $A_2$ for the unbarred galaxies remain well below the adopted threshold of $0.2$ (note that, for NB$_2$, we find a few radial bins with $A_2>A_2^\mathrm{thres}$, but the radial profile is clearly inconsistent with the presence of a bar). 

Previous works have classified bars into strong and weak ones, according to the maximum values achieved by $A_2$ in its radial profile ($A_2^{\rm max}$), with a general consensus that strong bars have  $A_{2}^{\rm max}> 0.3-0.4$ \citep{Algorry_2017, Fragkoudi2020, RosasGuevara_2022}. Our four bars enter the category of strong bars, as $A_2^{\rm max}$ are well above this threshold, as shown in Table \ref{tab:barsProps}.

\begin{table} 
    \centering
    \begin{tabular}{lcccccc}
        \hline
        Galaxy & $A_2^{\rm max}$ & $t_{\rm bar}$ & $R_{\rm bar}$ &  $\Omega_{\rm bar}$ & $R_{\rm CR}$ & $\mathcal{R}$ \\
        
        & - & Gyr & kpc & $\rm km \, s^{-1} \, kpc^{-1}$ & kpc & - \\
        \hline
        B$_1$ &0.58 & 3.74  & 3.32 & 28.33 & 5.93 & 1.90 \\
        B$_2$ &0.53 & 4.49   & 4.50 & 33.28 & 5.35 & 1.18 \\
        B$_3$ &0.54 & 10.29 & 4.95 & 34.89 & 5.43 & 1.17 \\
        B$_4$ &0.41 & 6.82  & 3.17 & 32.27 & 5.68 & 1.93 \\
        \hline
    \end{tabular}
    \caption{Properties of the bars in the barred galaxies, at $z = 0$: maximum value achieved by $A_2$ in its radial profile ($A_2^{\rm max}$); time of bar formation ($t_{\rm bar}$); bar length ($R_{\rm bar}$); pattern speed ($\Omega_{\rm bar}$); corotation radius ($R_{\rm CR}$); parameter distinguishing 'slow' from 'fast' rotator ($\mathcal{R}$). The parameters $\Omega_{\rm bar}$ and $R_{\rm CR}$ are estimated by using the code from \citet{Dehnen_2023}.}
    \label{tab:barsProps}
\end{table}

The position angle of a bar, denoted as $\Phi_{\rm{b}}$, is determined based on the $b_2$ and $a_2$ coefficients of the Fourier decomposition as   
$\Phi_{\rm{b}}(R) = 0.5 \, \rm{arctan(b_2/a_2)}$. 
In the case of a theoretically ideal bar, this phase should be constant along the bar and, as a result, it  can be used as an additional indicator of the presence of a bar. 
The lower panels of Fig.~\ref{fig:a2rad_phirad} show the bar phase as a function of projected radius, for our barred (left-hand panel) and unbarred (right-hand panel) galaxies, at $z=0$. 
In the case of galaxies with bars, the phase remains approximately constant up to $R\approx 4\,\mathrm{kpc}$, with some variation among galaxies. 
It is certainly not expected that, in the case of bars developed in cosmological simulations, the bar phase remains exactly constant. For this reason, we assume that small variations in the phase, i.e. those satisfying the condition $|\Delta \Phi_{\rm{b}}| < 0.15$, are consistent with a bar-like structure. In contrast to the barred galaxies, all the simulated unbarred systems show significant variations in the radial profile of the phase. 

Bar diagnostic quantities obtained from the Fourier decomposition can also be used to estimate the length of a bar. 
In this work, we assume that the bar length, $R_{\rm bar}$, is the maximum radius within which the bar phase is approximately constant (i.e. the innermost radius where $|\Delta \Phi_{\rm{b}}|$ becomes larger than $0.15$). The vertical lines in the left panels of  Fig.~\ref{fig:a2rad_phirad} show the bar lengths calculated in this way, which lie in the range $\sim 3-5\,\mathrm{kpc}$ (see Table~\ref{tab:barsProps}). 
It is worth noting that, at the outer end of the bars, the values of $A_2$ become lower than $A_{2,\mathrm{thresh}}$. In fact,  the radius at which this situation occurs provides an alternative way to estimate the bar length. 
Here, we adopt the constancy of the bar phase as our preferred criterion.   
While there are many different ways to define the bar length, \citep[e.g.][]{Athana_Misi2002,ScannapiecoAthanassoula_2012}, this is adequate for the purposes of this work and does not affect our results in any sensitive way.
The bar lengths we derive are consistent with the visual inspection of the bars in Fig.~\ref{fig:tot}.

\subsection{Pattern speed and corotation radius} \label{Sec:PatternSpeedCR}

The  pattern speed, $\Omega_{\mathrm{bar}}$, is another  key characteristic of a bar, which rotates as a solid body. $\Omega_{\mathrm{bar}}$ is the rotational frequency  of the bar  and, in contrast to the bar length and strength, is a purely dynamical property.
While we have the full kinematic information of all stars in each simulated galaxy, it is difficult to identify those stars that form the bar -- which would provide a clean measurement of the bar pattern speed -- as these are spatially mixed with bulge and disc material. Alternative ways of estimating this quantity have been proposed. The most direct approach is to apply the definition of the pattern speed using consecutive snapshots, i.e., calculate the time variation of the phase of the mode-2 Fourier transform within the bar. In order to get reliable results, adequate temporal resolution is needed so that changes in the position angle of the bar between different simulation outputs can be computed. However, this condition is hardly met in cosmological simulations.
An alternative approach is to use star particles near the outer end defined by the bar length, and calculate their rotational frequencies \citep{Marioni_2022}. The bar pattern speed can thus be estimated as the mean rotational frequency at the bar length. Another approach,  commonly used in observations but applicable to simulations as well,  is the Tremaine-Weinberg method \citep{Tremaine_Weinberg_84}. However, in the recent work of \cite{Dehnen_2023}, the authors argue that employing this method in simulations is not entirely suitable as it assumes a steady pattern and uses only one of the three velocity components. In order to avoid the limitations or previously used methods, \cite{Dehnen_2023} suggest an unbiased way of calculating the pattern speed of a galactic bar from a single simulation snapshot. 
We applied this method to measure the bar pattern speed of our simulated barred galaxies, and found values in the range $\sim 28-35\, \mathrm{km\ s^{-1}}$ at $z=0$, as shown in  Table~\ref{tab:barsProps}.

\begin{figure}
    \includegraphics[width=1\columnwidth]{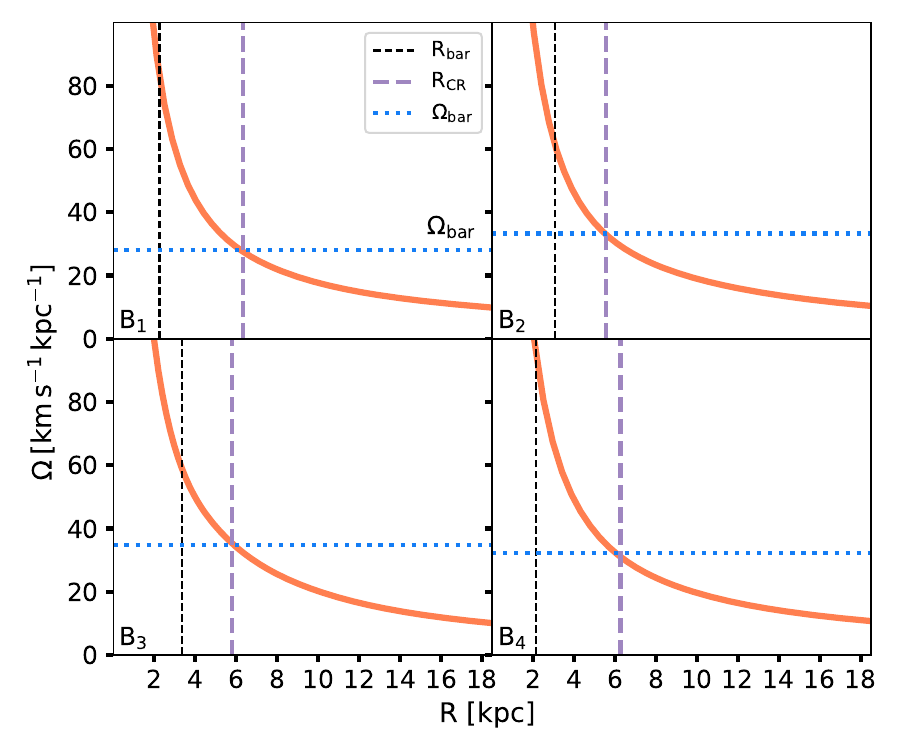}
    \caption{Angular frequency as a function of radius for the four barred galaxies, at $z=0$. The dotted horizontal lines indicate the corresponding  $\Omega_{\rm{bar}}$ values, obtained using the code described in \citep{Dehnen_2023}. We also show the length of the bars, $R_{\rm bar}$ (thin dashed vertical lines) and the corotation radii, $R_{\rm CR}$ (dashed vertical  lines).}
    \label{fig:phiRad_all_dehnen}
\end{figure}

The corotation radius, $R_\mathrm{CR}$, where the gravitational and centrifugal forces cancel out in the rest frame of the bar \citep{Guo2018}, is the radius at which the angular frequency of the galaxy matches that of the bar, $\Omega(R_{\mathrm{CR}})=\Omega_{\mathrm{bar}}$. 
Using the values obtained for $\Omega_{\mathrm{bar}}$ and estimating the angular frequency of the galaxy as $\Omega (r)= V_{{\rm circ}}(r)/r$ (with $V_{\mathrm{circ}}$ being the circular velocity of the galaxy), we calculate the corotation radii for our barred galaxies, at $z=0$. 
Fig.~\ref{fig:phiRad_all_dehnen} shows the radial profile of the angular frequency of each barred galaxy, at $z=0$, together with the value of $\Omega_{\mathrm{bar}}$ obtained through the \cite{Dehnen_2023} method, depicted by the horizontal blue dotted line. The intersection of these two lines determines the corotation radius, which are indicated by the vertical dashed purple line in this figure.
The values obtained for $R_{\mathrm{CR}}$ are in the range $\sim 5.3-6 \, \rm kpc$ (see Table~\ref{tab:barsProps}).
The corotation radius is an important quantity, as it can influence the stability of the bar. If the bar extends beyond $R_{\rm{CR}}$, the stars in the bar may have a higher angular velocity than the rotational velocity of disc  stars at that distance, which can lead to the bar weakening or breaking up. As shown in Fig.~\ref{fig:phiRad_all_dehnen}, $R_{\rm{CR}}$ is larger than  $R_{\rm{bar}}$ (identified by vertical dashed black lines), in all four cases.

Bars can also be classified into fast or slow rotators, using the parameter $\mathcal{R}$, defined as
the ratio between the corotation radius and the bar length, i.e., $\mathcal{R}={R_{\rm{CR}}}/{R_{\rm{bar}}}$.
Theoretical arguments \citep{Contopoulos1980} require $\mathcal{R}>1$, and there is a predisposition that $\mathcal{R} \gtrsim 1$ \citep{Sellwood&Wilkinson1993}. Bars with $1 \leq \mathcal{R} \leq 1.4$ are classified as `fast' bars, those for which corotation is not far beyond the bar’s end; those with $\mathcal{R} > 1.4$ are `slow bars' \citep{Debattista_2000}. 
 At $z=0$, our simulated bars have $\mathcal{R}$ values in the range $\sim 1.09-1.79\, \rm kpc$ (see Table~\ref{tab:barsProps}), 
 which set half of them into the slow rotator category and the other half into the fast rotator category.
 
\section{Evolution of bar properties} \label{Sec:EvolBarProp}

\subsection{Formation time of the bars} \label{Sec:tbar}

While the formation of a bar is not instantaneous, it is useful to define a formation time ($t_\mathrm{bar}$)  to identify when a bar can be considered fully formed. This allows us to study the properties of the galaxy that hosts the bar and its disc prior to $t_\mathrm{bar}$,  as well as  the evolution of the bar after its formation.
The most common way to estimate $t_\mathrm{bar}$ is through  the analysis of the time evolution of  $A_2^\mathrm{max}$ -- the maximum of the radial profile of the  $A_2$ parameter  (see Section~\ref{Sec:IdentificationBars})  -- and adopting a criterion to determine the presence or absence of a bar at each time. For consistency with our $z=0$ analysis, we assume that a galaxy has a bar if, at any time, $A_2^\mathrm{max} > A_2^\mathrm{thres}$ with $A_2^\mathrm{thres}=0.2$. Thus, the formation time is defined as the moment at which $A_2^{\rm max}$ surpasses $A_2^{\rm thres}$ and continues to grow, maintaining values above $A_2^{\rm thres}$ until $z=0$. This definition is adequate, with the only caution at those times where significant noise is present, which might occur at early times when galaxies are still in a first formation phase and the spatial distribution is highly asymmetric.

\begin{figure}
    \includegraphics[width=1\columnwidth]{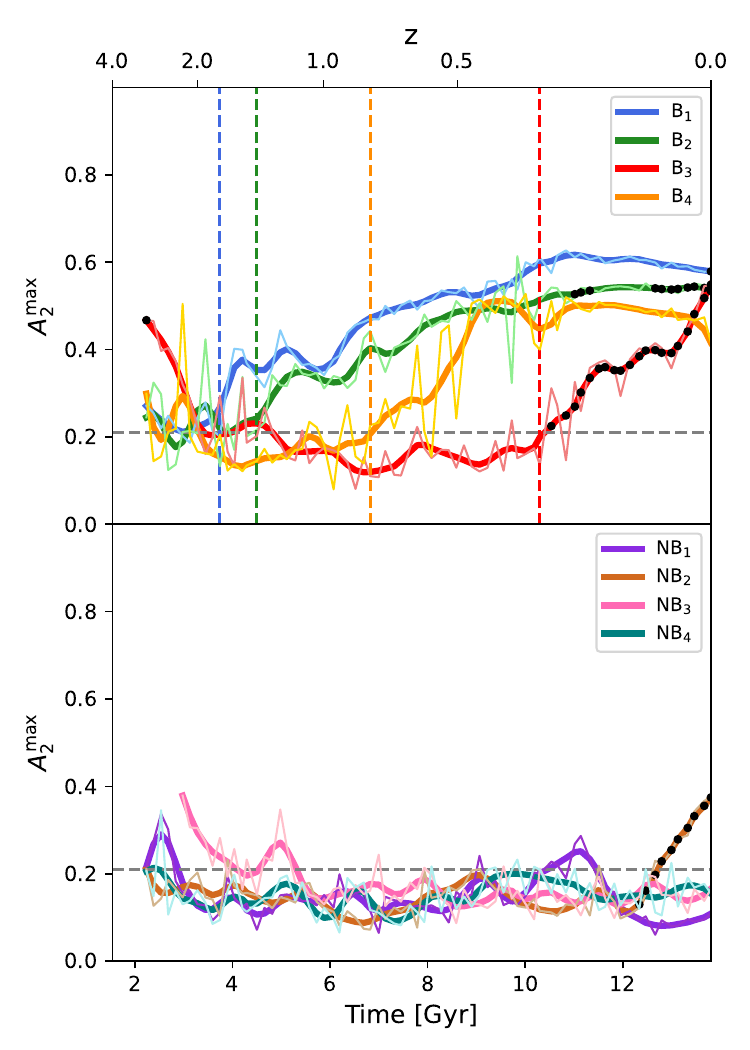}
    \caption{Time evolution of the maximum value of $A_2^{\rm max}$ for galaxies with a bar (upper panel) and galaxies without a bar (lower panel). The horizontal dashed line is the limit to determine the presence or absence of a bar in the galaxies ($A_2^{\rm thres}=0.2$). The vertical dashed lines correspond to the formation times of the bars ($t_{\rm bar}$) present in each galaxy. 
    This moment is defined as the point at which $A_2^{\rm max}$ surpasses $A_2^{\rm thres}$ and continues to grow, maintaining values above $A_2^{\rm thres}$ until $z=0$. The circles positioned over the curves denote the snapshots when the galaxies are identified as satellites. All the barred galaxies in our sample develop their bars while being central galaxies.
    }
    \label{fig:a2t}
\end{figure}

Fig.~\ref{fig:a2t} shows the time evolution of the $A_2^\mathrm{max}$ parameter for the simulated barred  (upper panel) and unbarred (lower panel) galaxies of our sample.
The corresponding  formation times of the bars  are indicated as vertical dashed lines in the figure. It is interesting to note that, even though we have only four galaxies in our sample, the formation times are diverse, with two systems forming their bars quite early, and the other two exhibiting a more recent formation (see Table~\ref{tab:barsProps}).
The black dots in this figure represent the times where  galaxies are identified as satellites of a more massive, central halo (thinner curves illustrate the raw computation, while the thicker curves result from a smoothing process employing a Butterworth filter). These galaxies  inhabit denser environments compared to the rest of the sample; we explore possible environmental effects in Section~\ref{Sec5}.

It is also worth noting that all simulated bars, and particularly those formed early on, have been able to survive until the present time. 
From the lower panel of this figure we can observe that, in general, the unbarred galaxies in our sample keep $A_2^\mathrm{max}$ values below the adopted threshold at all times, consistent with a lack of a bar, with the exception of  NB$_2$. As we have seen (Fig.~\ref{fig:a2rad_phirad}), the radial profiles of $A_2$  and $\Phi$ at $z=0$ for this galaxy is inconsistent with the presence of a bar, although by examining these radial profiles for earlier times it seems that NB$_2$ is starting to develop an asymmetry in the very inner regions, which could lead to the formation of a bar.

\subsection{Evolution of the bar length and 
pattern speed} \label{Sec:EvolBarLengthPatternSpeed}

\begin{figure}
    \includegraphics[width=1\columnwidth]{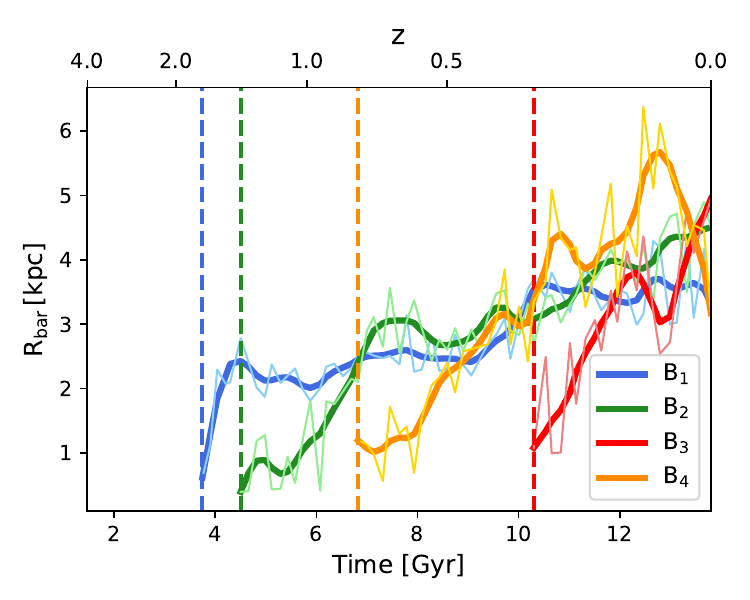}
    \includegraphics[width=1\columnwidth]{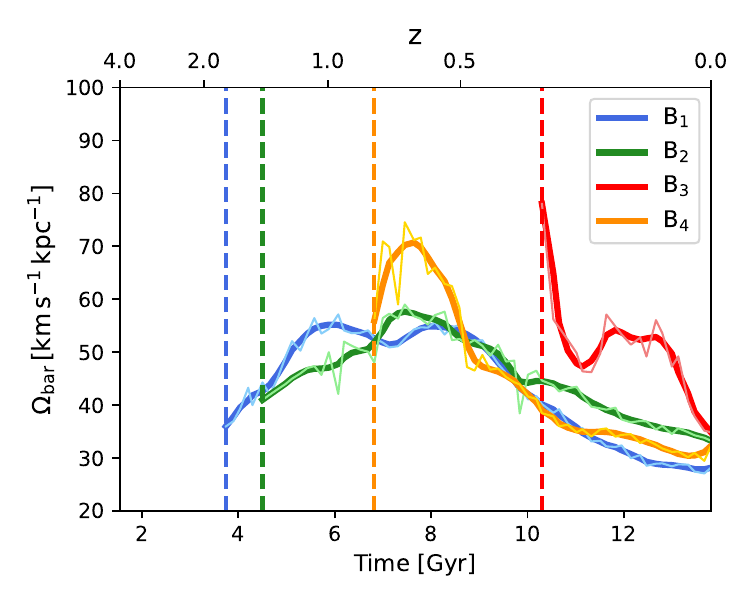}
    \caption{Time evolution  of the bar length, $R_{\rm bar}$ (upper panel) and  pattern speed, $\Omega_{\rm bar}$ (lower panel) 
    for the simulated barred galaxies, starting at the formation time of each bar, $t_{\rm bar}$ (vertical dashed lines). }
    \label{fig:bar_evolution}
\end{figure}

In this section we investigate the time evolution of the bar properties for the simulated  barred galaxies.   Fig.~\ref{fig:bar_evolution} shows the evolution of the bar length and the pattern speed of the bar as a function of time; starting at the formation time of each bar ($t_{\rm bar}$).
In all cases, the bar length  (upper panel) increases with time, varying from $\sim 1~\mathrm{kpc}$ at early times and reaching typical values of $\sim 4~\mathrm{kpc}$ at the present day, regardless of their formation time.  The galaxies whose bars formed early on, B$_1$ and B$_2$, show a smoother evolution of the bar length compared to B$_3$ and B$_4$.  

\begin{figure}
    \centering
    \includegraphics[width=1\columnwidth]{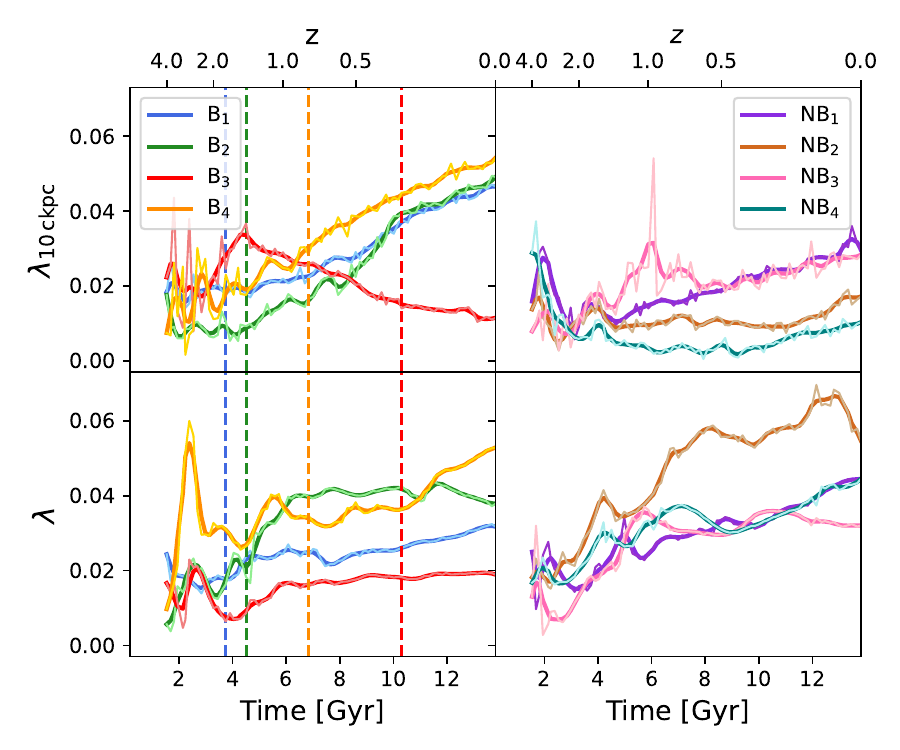}
    \caption{\textit{Upper panels}: Time evolution of the spin of the dark matter halo considering the dark matter contained within $10\,{\rm ckpc}$, $\lambda_{\rm 10\,ckpc}$ for barred galaxies (\textit{left}) and for unbarred galaxies (\textit{right}). \textit{Lower panels}: Same as upper panels but considering the dark matter within the virial radius; in this case, we refer to the spin parameter simply as $\lambda$.The vertical dashed
    lines in the left-hand panels correspond to the formation times of the bars ($t_{\rm bar}$).}
    \label{fig:HaloSpin}
\end{figure}

\begin{figure*}
    \includegraphics[width=2\columnwidth]{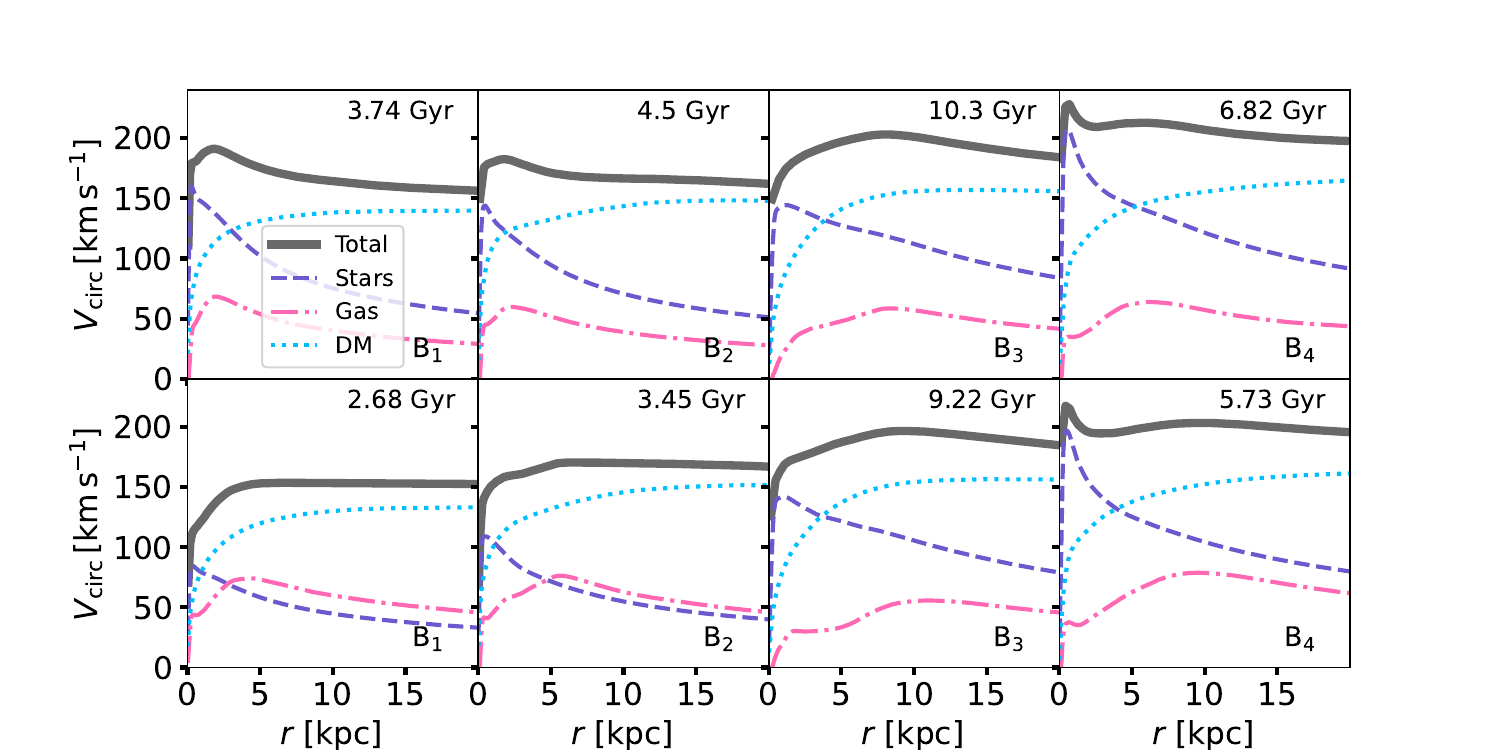}
    \caption{Circular velocity curves (solid lines) for barred galaxies at $t_{\rm{bar}}$ (\textit{upper panels}) and at $\sim 1\,{\rm Gyr}$ before the bar formation time (\textit{lower panels}), as a function of radius $r$. The contributions of the dark matter, stellar and gaseous components are identified by dotted, dashed and dashed-dotted lines, respectively, as indicated in the legend.}
    \label{fig:Vcirc_past}
\end{figure*}

Bars that have formed more recently (B$_3$ an B$_4$) achieved the highest maximum values of  $\Omega_\mathrm{bar}$ ($\sim 70\,{\rm km}\,{\rm s}^{-1}$). 
The values of $R_{\rm bar}$ and $\Omega_{\rm bar}$ achieved at $z=0$ are so similar (see Table~\ref{tab:barsProps}) that we cannot conclude from this small sample that shorter bars rotate faster, and vice versa, as expected
\citep{Cuomo_2020}.
In general, the pattern speed of the  bars (lower panel of Fig.~\ref{fig:bar_evolution}) increases right after the formation of the bar for a short period and decreases thereafter, accompanied by an increase of the bar length.  

The decrease in the bar pattern speed is consistent with expectations that angular momentum keeps being transferred outwards to the disc and halo after the bar has formed, a scenario suggested by gas-less simulations focusing on non-rotating, isolated dark matter haloes \citep[][and references therein]{Athanassoula_2014}.
This transfer occurs through the interactions of disc and halo particles trapped at lower resonances. Bars experiencing a loss of angular momentum undergo a slowdown and strengthening due to the transition of stellar orbits within the bar towards a more radial configuration. The reduction of the bar's pattern speed facilitates the sweeping of orbital resonances across the majority of the dark matter halo phase space, leading to the widespread distribution of angular momentum among the halo particles. Consequently, the angular momentum of the halo increases over time
\citep{Dubinski_2009}. The effectiveness of this angular momentum transfer depends on the amount of near-resonant material and its dynamical temperature \citep{Athanassoula_2003}. 
Despite the typically higher temperature of the halo's material, bars within substantial haloes exhibit greater strength compared to bars in less massive haloes due to their significant contribution of material in relation to the outer disc.

More recently, utilising models of stellar disc galaxies within dark matter haloes, it has been demonstrated that the dynamic and secular evolution of bars is influenced by various properties of their parent dark matter halo. These properties include the cosmological spin parameter (a dimensionless representation of angular momentum), halo shape, and density distribution
\citep{SahaNaab_2013, Long_2014, Collier_2018, LiShlosman_2023, Joshi_2024}. 
The halo spin at the virial radius, $R_{\rm vir}$, is given by $\lambda=J_{\rm h}/(\sqrt{2}\,M_{\rm vir}\,R_{\rm vir}\,V_{\rm circ}(R_{\rm vir}))$, where $J_{\rm h}$ is the angular momentum of the dark matter halo and $V_{\rm circ}(R_{\rm vir})$ is the circular velocity at $R_{\rm vir}$ \citep{Bullock_2001}. 
The expected range of cosmological spin values spans $\lambda \sim 0-0.09$, with  $\bar{\lambda}\sim 0.035-0.04$ representing the mean value of its lognormal distribution. In these simulations of isolated disc-halo systems, the bar evolution is clearly divided into two phases: an initial dynamical phase from bar instability to the first vertical buckling instability (a process that breaks the symmetry with respect to the disc equatorial plane resulting in the vertical thickening of the bar and the formation of a boxy/peanut-shaped bulge), and a subsequent phase of secular growth of the stellar bar.
Key findings from these investigations reveal that the time-scale of bar instability shortens with growing $\lambda$, and that the effectiveness of angular momentum absorption by haloes diminishes with increasing $\lambda$ values. As a consequence of the latter aspect, the bar pattern speed experiences a sudden decrease during buckling for models with low spin ($\lambda \lesssim 0.03$), while it remains rather constant for models with higher spin values ($\lambda \gtrsim 0.03$). As a result, outcomes for low spin values are consistent with those observed in non-rotating haloes, as discussed previously.  
It is noteworthy that varying either the halo shape or the dark matter density affects the evolution of the bar along $\lambda$-sequence \citep{Collier_2018, LiShlosman_2023}.

While the studies described previously suggest that faster-spinning haloes reduces the amplitude of stellar bars during their secular evolution, this finding is contradicted by the recent analysis conducted by \cite{KatariaShen_2022}.
They observe that bars continue to grow during secular evolution regardless of the halo spin. The discrepancy arises because these authors examine models with increasing spin achieved by augmenting the angular momentum in the central regions ($<30\,{\rm kpc}$) of the galaxy, as most of the disc–halo angular momentum interaction occurs within the bar region\footnote{Because the halo extends across a substantial radial scale in comparison to the disc, different radial distributions of internal halo angular momenta can give rise to the same total spin.}.
Consequently, their results demonstrate that although the disc-to-halo transfer of angular momentum decreases with increasing spin, strong bars can still exchange angular momentum and grow through torquing down due to the dynamical friction of the haloes.

Our sample of galaxies with and without bars is derived from a cosmological hydrodynamical simulation, placing them within dark matter haloes that exhibit varying values of key properties such as mass, concentration, and spin.
We now examine the influence of the dark matter halo spin on determining the presence or absence of a bar in our chosen galaxy sample. Fig.~\ref{fig:HaloSpin} considers the evolution of the spin parameter for both barred and unbarred galaxies. The spin parameter for dark matter particles within $10\,{\rm ckpc}$, $\lambda_{\rm 10\,{\rm ckpc}}$, is presented in the upper panels, while the lower panels depict the spin parameter estimated from dark matter particles within $R_{\rm vir}$, denoted simply as $\lambda$. The former choice involves estimating the spin approximately at twice the corotation radius of barred galaxies (see values of $R_{\rm CR}$ in Table~\ref{tab:barsProps}), and is based on the fact that bar formation is more likely to be affected by the inner halo angular momentum close to the bar region    \citep{KatariaShen_2022}. 
In this figure, it is evident that the initial values of the dark matter halo spin, $\lambda_{10\,{\rm ckpc}}$ and $\lambda$ are below $\sim 0.03$ for both barred and unbarred galaxies. The initial values are determined at $\sim 2.5\,{\rm Gyr}$ into the galaxy's evolution, a point when the galaxy's disc is well-defined, as confirmed through visual inspection of the time evolution of the stellar mass density in three spatial projections.

Focusing on the evolution of $\lambda_{10\,{\rm ckpc}}$, we observe that their values increase monotonically with time for barred galaxies (with the exception of B$_3$) and present more variable trends for unbarred galaxies—either increasing, remaining rather flat, or also decreasing during some periods. The $z=0$ spin values achieved by barred galaxies exceed $\sim 0.04$, larger than those reached by unbarred galaxies, which do not exceed this value in any case.
The general trend indicates that bars form as a result of substantial angular momentum transfer from the disc to the halo, a process not occurring in unbarred galaxies. While this aligns with results from simulations of isolated disc-halo systems, the crucial aspect to emphasise from our results is that the initial value of $\lambda_{10\,{\rm ckpc}}$ is similar in all cases and close to the low spin values that promote bar formation according to other studies \citep[e.g.][]{Collier_2018, LiShlosman_2023}. However, despite that, some galaxies do not develop a bar. 
Thus, it seems that the spin of the dark matter halo cannot be regarded as the primary and only  factor influencing bar evolution, and additional modulation by other halo properties (mass, shape, concentration) as well as the influence of external processes might also play a role in the development of disc instabilities and bar growth, as envisaged by previous studies.
In this context, it is noteworthy that the unique behavior in the spin evolution of galaxy B$_3$ could be elucidated by the fact that this galaxy becomes a satellite as soon as it develops a bar, and its preceding dynamical history might be influenced by the high density environment in which it is moving prior to being accreted by a larger halo (see Fig.~\ref{fig:Environment} and Section~\ref{Sec:ExternalTriggers}).

On the contrary, the trajectory traced by the overall spin parameter $\lambda$ (depicted in the lower panels of Fig.~\ref{fig:HaloSpin}) deviates significantly from the pattern illustrated by $\lambda_{10\,{\rm ckpc}}$: it remains relatively constant for barred galaxies, staying below $\sim 0.04-0.05$ at $z=0$, while steadily increasing for unbarred galaxies, reaching values exceeding $\sim 0.04$. Clearly, the redistribution of angular momentum among dark matter particles exhibits notable distinctions between barred and unbarred galaxies. Examining the temporal evolution of the radial distribution of the spin parameter could provide crucial insights into unraveling the underlying cause of bar formation in disc galaxies. A more detailed analysis of such aspects is intended for future investigations.

\subsection{Evolution of the central mass concentration} \label{Sec:StellartoDmMassRatio}

\begin{figure}
    \centering
    \includegraphics[width=1\columnwidth]{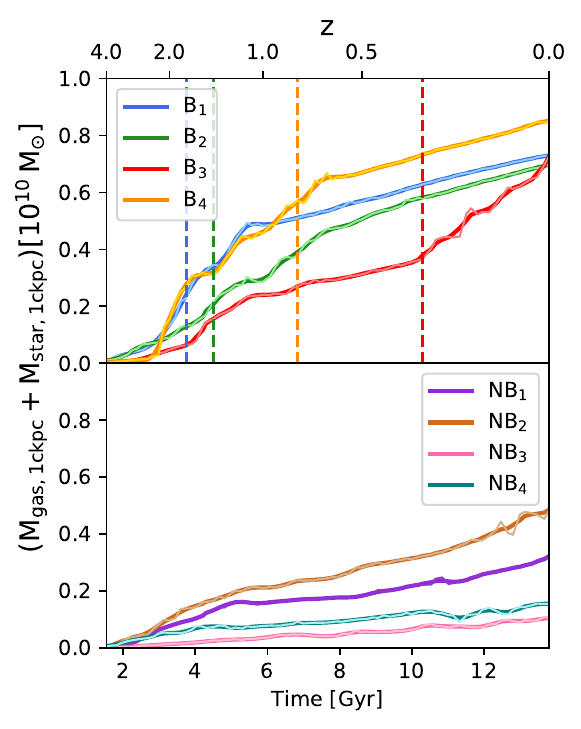}
    \caption{Time evolution of the central mass concentration (CMC), defined as the sum of gas and stars within $1\,{\rm ckpc}$ for barred galaxies ({\em upper panel}) and unbarred galaxies ({\em lower panel}). The vertical dashed lines indicate the formation time of the bars ($t_{\rm bar}$).}
    \label{fig:CMC1ckpc}
\end{figure}

As we have shown in Fig.~\ref{fig:Vcirc}, the simulated unbarred galaxies have, at $z=0$, smoother rotation curves compared to galaxies with bars, which exhibit a peak in the inner regions dominated by the stellar component.
Here, we show that this characteristic of the galaxies that form bars originates at early times, even prior to their development.  Fig.~\ref{fig:Vcirc_past} shows circular velocity curves for the barred galaxies, including the total circular velocity, as well as the separate contribution of the dark matter, stellar and gaseous components.
We show, for each bar, the circular velocity curves at the formation time of the bars ($t_\mathrm{bar}$), as well as at approximately $1\,{\rm Gyr}$ before the bar formation time ($\sim t_\mathrm{bar} - 1~\mathrm{Gyr})$. In all cases, the central mass budget is predominantly governed by the stellar component, with its contribution surpassing that of gas and dark matter.
This behaviour is not originated as a consequence of the formation of the bar, as
it is already evident prior to the bar emergence. 
It is worth noting that the central stellar mass concentration becomes more pronounced when the bar forms, reaching its maximum at $z=0$ (Fig.~\ref{fig:Vcirc}).
This aligns with findings reported by \cite{Dubinski_2009} in their examination of bar instability in galactic models featuring an exponential disc and a cuspy dark matter halo. 
Both stars and gas within the corotation radius experience a loss of angular momentum, causing them to move inward gradually. Consequently, the formation of the bar also results in an enhanced central concentration in the gas component.

The central mass concentration (CMC) is defined as the total baryonic mass (comprising stars and gas) within the central region \cite[e.g.][]{Athanassoula_2013, Seo_2019}. To explore the relationship between the development of this CMC and bar formation, we examine the temporal evolution of the CMC for both barred and unbarred galaxies, 
considering the baryonic mass contained within $1\,{\rm ckpc}$ (comoving kpc). This is shown in our Fig.~\ref{fig:CMC1ckpc}. The CMC increases monotonically with time for both barred and unbarred galaxies, with a higher rate of increase observed for the former, resulting in larger CMC values at $z=0$. Barred galaxies generally exhibit a higher CMC throughout their evolution, aligning with our previous findings based on circular velocity curves. For all barred galaxies except B$_3$, the rate of increase tends to decrease after bar formation, while the opposite is true for B$_3$. As discussed in the previous section and elaborated on later, B$_3$, residing in a high-density environment and becoming satellite after forming its bar, might experience close encounters triggering gas inflows that contribute to a more abrupt increase in its CMC.

These patterns are in contrast to findings that suggest a scenario where the formation of a high mass concentration, such as a massive and compact bulge, hinders bar development \citep{Kataria_2018, SahaElmegreen_2018} or prevents the formation of strong and long bars \citep{Jang_2023}. These studies have utilized N-body simulations incorporating dynamic halo and bulge components. When accounting for the presence of gas, the situation becomes more intricate as gas, being dissipative, is highly responsive and can alter the density distribution of the entire disc. Once bars are formed, they can induce gas inflows \cite{ReganTeuben_2004, Berentzen_2007, Shin_2017}.
In numerical simulations of discs embedded in rigid haloes with gas dynamics considered, \cite{Bournaud_2005} find that a growing CMC does not completely dissolve a bar. However, the gravitational torques exerted by gas on the bar cause orbits to become rounder, significantly weakening and dissolving it in $\sim 2\,{\rm Gyr}$. According to \cite{Berentzen_2007}, the bar is not destroyed, but its amplitude decreases due to vertical buckling and regrows after this process concludes. In gas-rich models, vertical asymmetry (buckling) of the bar is dampened by the forming CMC, which heats up the central kiloparsec in the stellar disc on a dynamical timescale but also puffs it up. Thus, the degree of stellar thickening is practically independent of the gas fraction in the disc.

Considering simulations of disc galaxies that include a gaseous disc component undergoing star formation and feedback, \cite{Athanassoula_2013} demonstrate that bars, while not completely destroyed, weaken more significantly in galaxies with larger fraction of gas since the CMC component is more massive in simulations with more
gas. More recently, \cite{Seo_2019} conducted high-resolution simulations of Milky Way-sized, isolated disc galaxies comprising a live halo, and stellar and gaseous discs. These simulations incorporate radiative heating and cooling, star formation and feedback, and assessed the effect of gas on bar formation in models with a cold and a warm disc that differ in radial velocity dispersions. They find  that a bar forms earlier and more strongly in the cold discs with larger fraction of gas, while gas progressively delays bar formation in the warm discs due to the larger value of the Toomre parameter, $Q_{\rm T}$.
The CMC grows faster for a stronger bar, resulting in a faster decay of the bar strength after reaching a peak because the CMC excites stellar motion in the vertical direction. Therefore, overall, the bar strength in the late phase of the disc evolution is inversely proportional to the gas fraction, independent of $Q_{\rm T}$.

\begin{figure}
    \includegraphics[width=1\columnwidth]{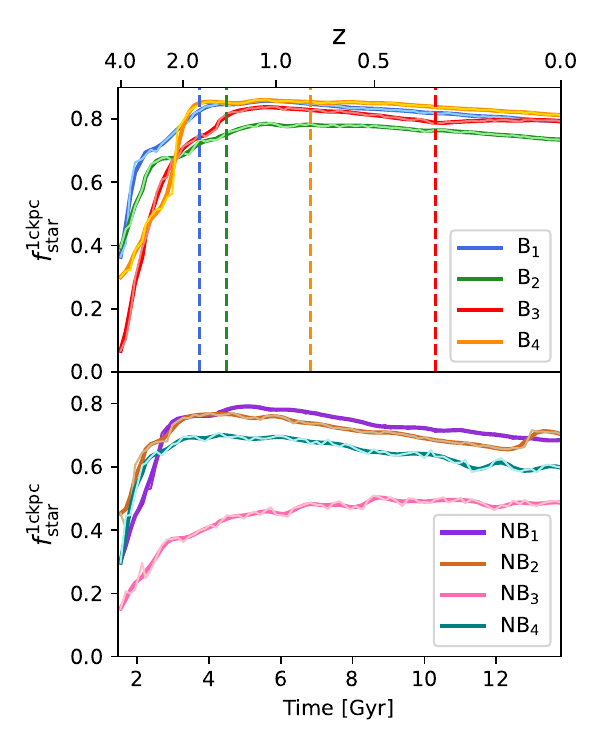}
    \caption{Time evolution of the mass ratio $f_\mathrm{star}^{1\mathrm{ckpc}}$
    between the stellar component and the sum of dark matter mass and stellar mass within $1\,{\rm ckpc}$, for barred galaxies (\textit{upper panel}) and for unbarred galaxies (\textit{lower panel}). The vertical dashed lines indicate the formation time of the bars ($t_{\rm bar}$).}
    \label{fig:ms_md}
\end{figure}

\begin{figure}
    \includegraphics[width=1\columnwidth]{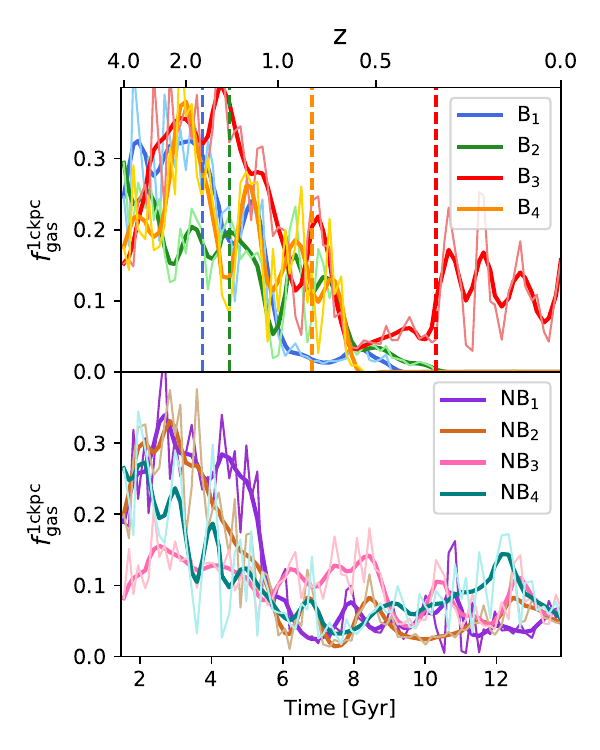}
    \caption{Time evolution of the mass ratio $f_\mathrm{gas}^{1\mathrm{ckpc}}$
    between the gaseous component and the sum of dark matter mass and gas mass within $1\,{\rm ckpc}$, for barred galaxies (\textit{upper panel}) and for barred galaxies (\textit{lower panel}). The vertical dashed lines indicate the formation time of the bars ($t_{\rm bar}$).}
    \label{fig:mg_md}
\end{figure}

To understand the role of the CMC in our galaxy sample and the respective contributions of stars and gas, we evaluate the relative proportions of these components with respect to dark matter within the inner regions. As previously observed in the circular velocity curves presented in Fig.~\ref{fig:Vcirc}, there are systematic differences between barred and unbarred galaxies in the mass distribution of baryons and dark matter.
We quantify this using the ratio $f_\mathrm{star}^{1\mathrm{ckpc}}\equiv M_\mathrm{star}^{1\mathrm{ckpc}}/(M_\mathrm{star}^{1\mathrm{ckpc}}+M_\mathrm{DM}^{1\mathrm{ckpc}})$, where $M_\mathrm{star}^{1\mathrm{ckpc}}$ and $M_\mathrm{DM}^{1\mathrm{ckpc}}$ refer to the stellar and dark matter masses in the inner $1~\mathrm{ckpc}$,  and the ratio
$f_\mathrm{gas}^{1\mathrm{ckpc}}\equiv M_\mathrm{gas}^{1\mathrm{ckpc}}/(M_\mathrm{gas}^{1\mathrm{ckpc}}+M_\mathrm{DM}^{1\mathrm{ckpc}})$, where $M_\mathrm{gas}^{1\mathrm{ckpc}}$ represents the mass of the gaseous component in the inner $1~\mathrm{ckpc}$.
Figs.~\ref{fig:ms_md} and~\ref{fig:mg_md} show, respectively, the evolution of these ratios for the barred (upper panel) and unbarred (lower panel) galaxies, along with the formation time of the bars.

The overall trend of $f_\mathrm{star}^{1\mathrm{ckpc}}$ is similar for both barred and unbarred galaxies, indicating a rapid and significant increase in the stellar mass relative to dark matter for both types of galaxies. An exception is NB$_3$, where the increase in the ratio of stellar to dark matter mass is smoother and sustained over longer time periods. These behaviours are, of course, entirely consistent with the shape of the velocity curves in the inner regions. The main difference between barred and unbarred galaxies is that the former reach values of $f_\mathrm{star}^{1\mathrm{ckpc}} \gtrsim 0.7$ after its rapid rise during the early stages of galaxy evolution, which is larger than those attained by unbarred galaxies spanning a wider range although initial values are within the same range for both types of galaxies ($f_\mathrm{star}^{1\mathrm{ckpc}}\lesssim 0.4$). 
On the other hand, Fig.~\ref{fig:mg_md} shows that, in general, galaxies with bars tend to possess larger gas reservoirs compared to their non-barred counterparts, especially before the formation of bars.
This aligns with observations indicating that gas-rich galaxies exhibit longer and stronger bars compared to HI gas-deficient galaxies at a constant stellar mass \citep{Zhou_2021}.

It is noteworthy that the proportions of stars and gas in relation to dark matter, along with the circular velocity curves for each component, indicate that unbarred galaxies are characterised by a higher amount of dark matter.
This result is consistent with those of \cite{Athanassoula_2013} and \cite{Reddish_2022} who find that the presence of a massive spheroid can prevent the formation of bars.  
Bars do not form in systems where dark matter dominates in the inner regions. 
Our results are also in line with the work of  \cite{Bland-Hawthorn_2023}, based on a hydrodynamical simulations of a halo-bulge-disc systems, who argue that the dominance of inner baryons over the dark matter component in early massive galaxies is a key factor in producing galactic bars.

\begin{figure}
    \includegraphics[width=1\columnwidth]{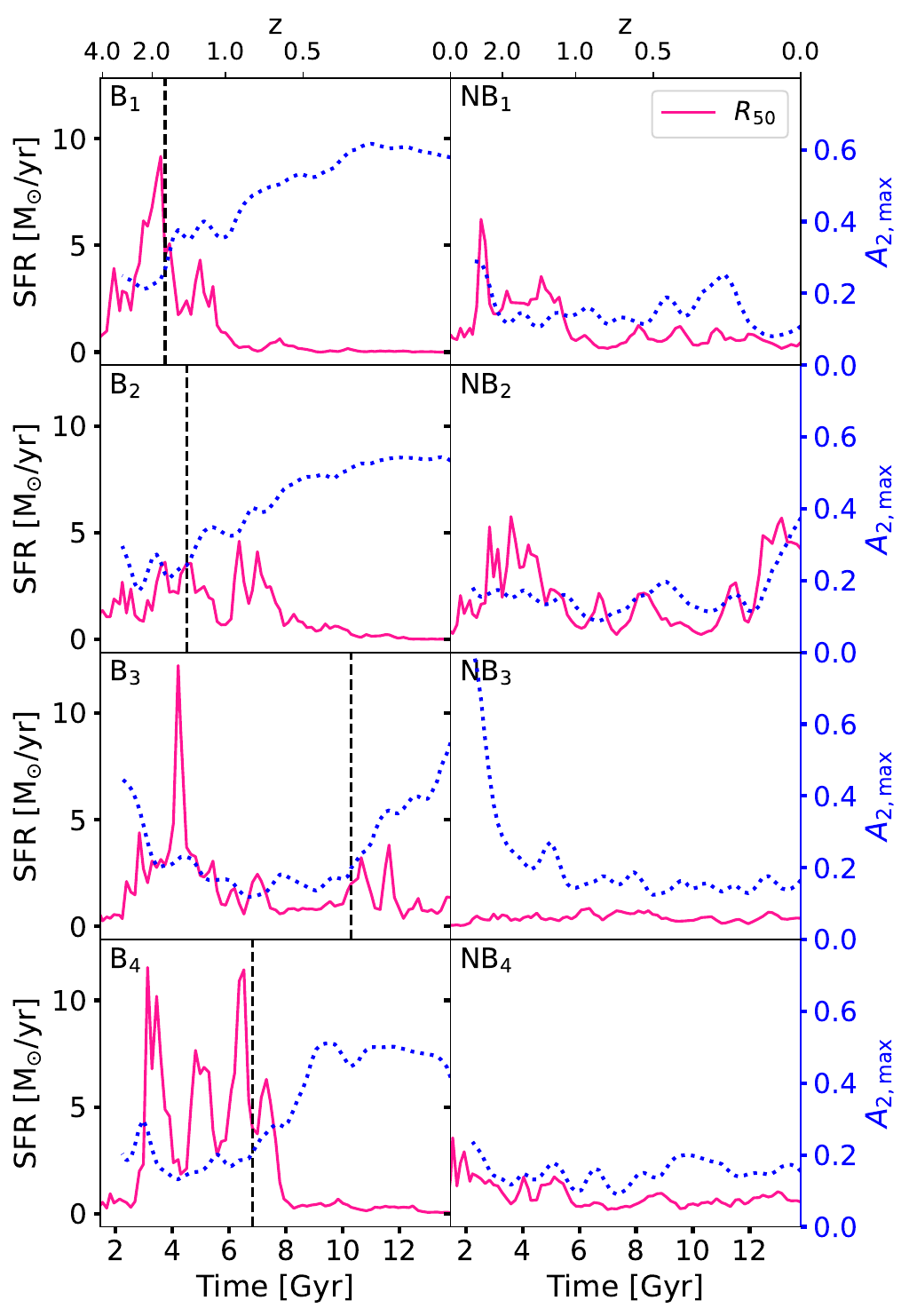}
    \caption{Time evolution of the star formation rate within the stellar half-mass radius (solid lines), for barred (left panels) and unbarred (right panels) galaxies.  The vertical dashed lines indicate the formation times of bars, $t_{\rm bar}$. The  dotted lines correspond to the right y-axis and show the time evolution of $A_{2, \rm max}$.}
    \label{fig:sfr}
\end{figure}

Our findings highlight the occurrence of substantial gas inflows during the formation of bars, resulting in elevated levels of star formation not only in the very inner regions, but also within the stellar half-mass radius of the barred galaxies, as illustrated in Fig.~\ref{fig:sfr}. Galaxies possessing bars display more prominent star formation bursts compared to unbarred galaxies, which exhibit smoother rates throughout their evolution, except for galaxy NB$_2$, which shows evidence of non-axisymmetry at $z=0$ (see Fig.~\ref{fig:tot}). The more pronounced bursts of star formation in barred galaxies contribute to lower fractions of cold gas in the inner regions at lower redshifts, consistent with the circular velocity curves presented in Fig.~\ref{fig:Vcirc}. An exception is galaxy B$_3$, characterised by particular features in relation to inhabiting a richer environment and becoming a satellite of a larger system, where bar formation occurs later than in the other barred galaxies, and there has not been sufficient time to deplete the central gas, with potential interactions triggering new gas inflows.

In summarising the results gleaned from the analysis of evolutionary trends in various properties of barred and unbarred galaxies, it becomes apparent that both sets of galaxies initially are located within haloes characterised by comparable spin values and share similarities in their initial stellar content. However, the initial gas levels in the inner regions are higher for barred galaxies. The subsequent substantial increases in stellar mass and spin values exhibited by barred galaxies result from the ongoing disc instability and bar development, involving the transfer of angular momentum from the disc to the dark matter halo. The associated increments in gas inflows contribute to a larger stellar content in the centre, leading to a more significant increase in the CMC in barred galaxies. This latter aspect aligns with findings from simulations of isolated disc-halo systems, as mentioned earlier, indicating that a higher gas fraction leads to a larger growth of the CMC \citep[e.g.][]{Athanassoula_2013}. However, the  contradiction lies in the fact that strong bars can develop when there are large initial gas fractions, and consequently, in the presence of a high CMC.
 
Concerning the contrasting correlation with the initial gas fraction exhibited by our model galaxies, it is crucial to underscore that, unlike in idealised simulations, the amount, characteristics, and distribution of gas are contingent on various interrelated processes. These processes, including gas accretion (from the intergalactic medium, if it can permeate the halo), gas cooling, star formation, and feedback mechanisms, interact in a complex and non-trivial manner. Stellar feedback, in particular, induces substantial alterations in the gas distribution: following periods of star formation, supernova explosions rapidly release a significant amount of energy, injecting pressure and prompting the circulation of gas from the star-forming regions  to the outer parts of the galaxy. In the context of bar formation, where gas inflows and star formation appear to play a key role, it is imperative to acknowledge that feedback will inevitably occur and influence the amount and spatial distribution of gas in galaxies. The intricate interplay among these processes makes it exceedingly challenging to draw definitive conclusions about how gas affects the bar formation or how the bar impacts the gas component, particularly with the concurrent formation of a CMC. 
These aspects deserve a more in-depth analysis of the hydrodynamical connection between the inner disc of the galaxy, the formation of the bulge, and the inner disc's susceptibility to bar formation under specific conditions within a cosmological context where galaxies undergo continuous growth and evolution.

\section{Bar formation: instabilities and mergers} \label{Sec5}

\subsection{Disc instabilities} \label{Sec:DiscInstabilities}

\begin{figure}
    \includegraphics[width=1\columnwidth]{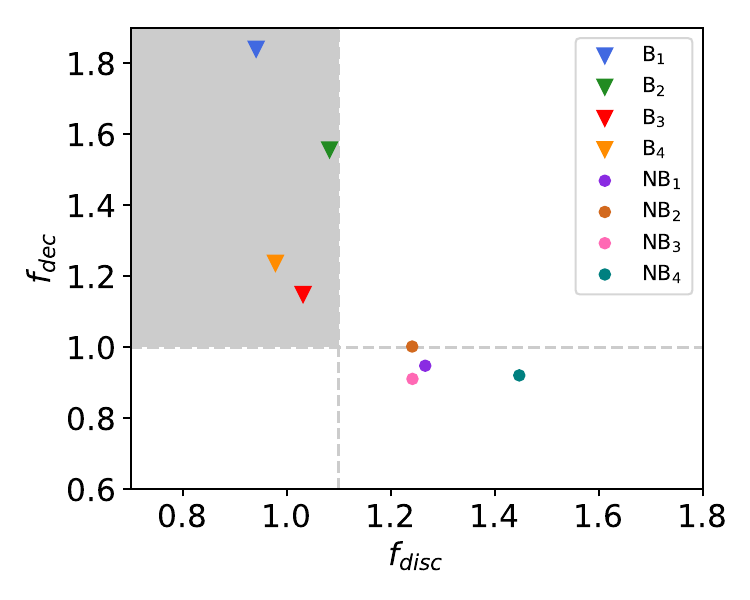}
    \caption{Disc instability parameters $f_{\rm{dec}}$ and $f_{\rm{disc}}$. The parameters are measured at the time formation of the bar, $t_{\rm{bar}}$, for the barred galaxies, and at $13.8 \, \rm{Gyr} \, (z=0)$ for the unbarred galaxies. Horizontal and vertical dashed lines indicate the corresponding thresholds: $f_{\rm{dec}}=1$ and $f_{\rm{disc}}=1.1$. The shaded area denotes the instability region.}
    \label{fig:fdec-fdisc}
\end{figure}

It is known that a favourable scenario for the formation of bars occurs when an instability is triggered in the stellar disc \citep{Sellwood_1993}. Several works have approached the study of instability mechanisms, distinguishing between `global'  or `local' types \citep{Toomre_1964,Efstathiou_1982}. 
In this Section, we analyse a number of proposed parameters to determine whether a system undergoes stellar disc instabilities, and test if they are able to predict the formation of bars in the simulations.

Following \cite{Algorry_2017}, we first calculate two of these parameters. The first one, referred to as $f_{\mathrm{disc}}$, measures the relative importance of the disc and the dark halo, and is defined as:
\begin{equation}
    f_{\mathrm{disc}} = \frac{V_{\rm circ}(R_{\rm{50}})}{\sqrt{GM_{\rm star}/R_{\rm{50}}}}
\end{equation}
where $R_{\rm{50}}$ is the stellar half-mass radius, $V_{\rm circ}(R_{\rm{50}})$ is the total circular velocity evaluated at the stellar half-mass radius, and $M_{\rm star}$ is the stellar mass of the galaxy. This parameter measures the \textit{local} gravitational importance of the disc and is an approximation of  the parameter $\epsilon_{\rm ELN}$ \citep{Efstathiou_1982} discussed in Sec.~\ref{Sec:Intro}. If $f_{\mathrm{disc}} < 1.1$, the disc is unstable to bar formation.

The second parameter, $f_{\mathrm{dec}}$, measures the \textit{global} gravitational importance of the disc. It is defined as:
\begin{equation}
    f_{\mathrm{dec}} = \frac{V_{\rm circ}(R_{\rm{50}})}{V_{\rm {circ},\rm{max}}^{\rm{dm}}}
\end{equation}
where $V_{{\rm circ},\rm{max}}^{\rm{dm}}$ is the maximun circular velocity of the dark matter halo. If $f_{\mathrm{dec}} > 1$,  the speed of rotational velocity of the disc is high with respect to the speed of the dark matter halo, and it is prone to the development of a bar. 

Fig. \ref{fig:fdec-fdisc} shows  $f_{\mathrm{disc}}$ versus $f_{\mathrm{dec}}$ for our barred  (inverted triangles) and  unbarred (full circles) galaxies. The parameters are measured at the time of formation of the bar ($t_{\rm{bar}}$) for the barred galaxies, and at $z=0$\footnote{The choice to evaluate the parameters at $z=0$ is somewhat  arbitrary (in \cite{Algorry_2017}, $z=0.5$ is chosen instead) but it should be adequate according to the discussion of our Section~\ref{Sec:StellartoDmMassRatio} and the results shown in the remaining of this Section.} for the unbarred galaxies. The grey area delimits the instability region, where $f_{\mathrm{disc}} < 1.1$ and  $f_{\mathrm{dec}} > 1$. All barred (unbarred) galaxies fall in the instability (stability) regions in this figure.

\begin{figure}
    \includegraphics[width=1\columnwidth]{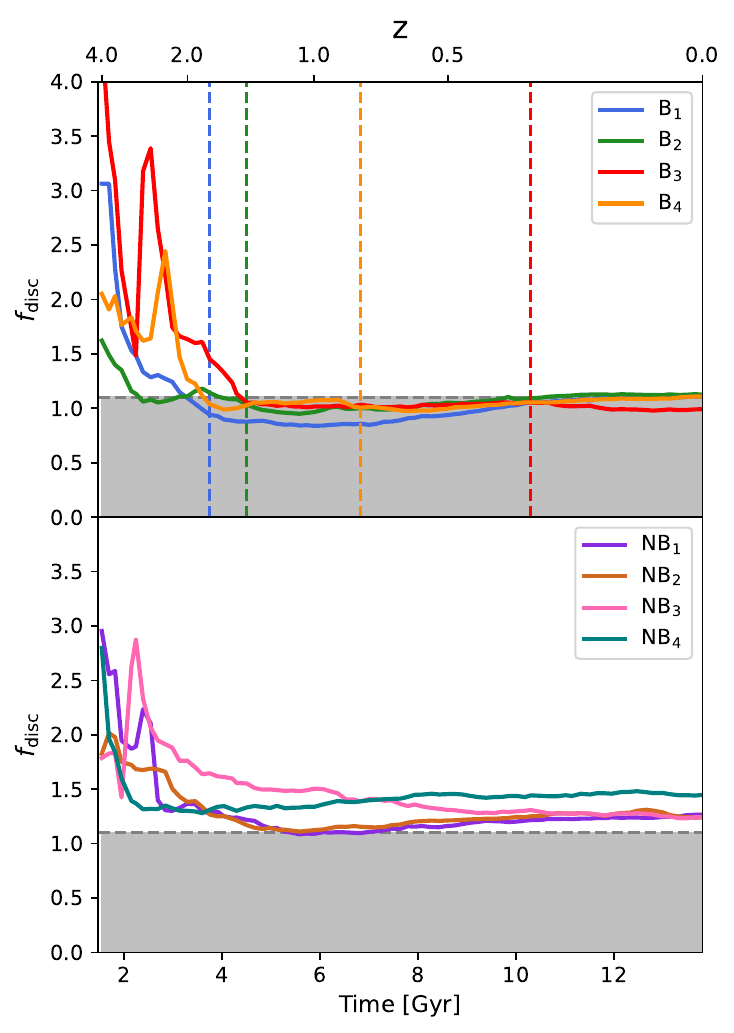}
    \caption{Time evolution of the parameter $f_{\rm{disc}}$, for the barred galaxies (\textit{upper panel}) and  the unbarred galaxies (\textit{lower panel}). The vertical dashed lines in the upper panel are the time formation of each bar, $t_{\rm bar}$, and the shaded area depicts the instability region: $f_{\rm{disc}}<1.1$.}
    \label{fig:fdisc-evol}
\end{figure}

\begin{figure}
    \includegraphics[width=1\columnwidth]{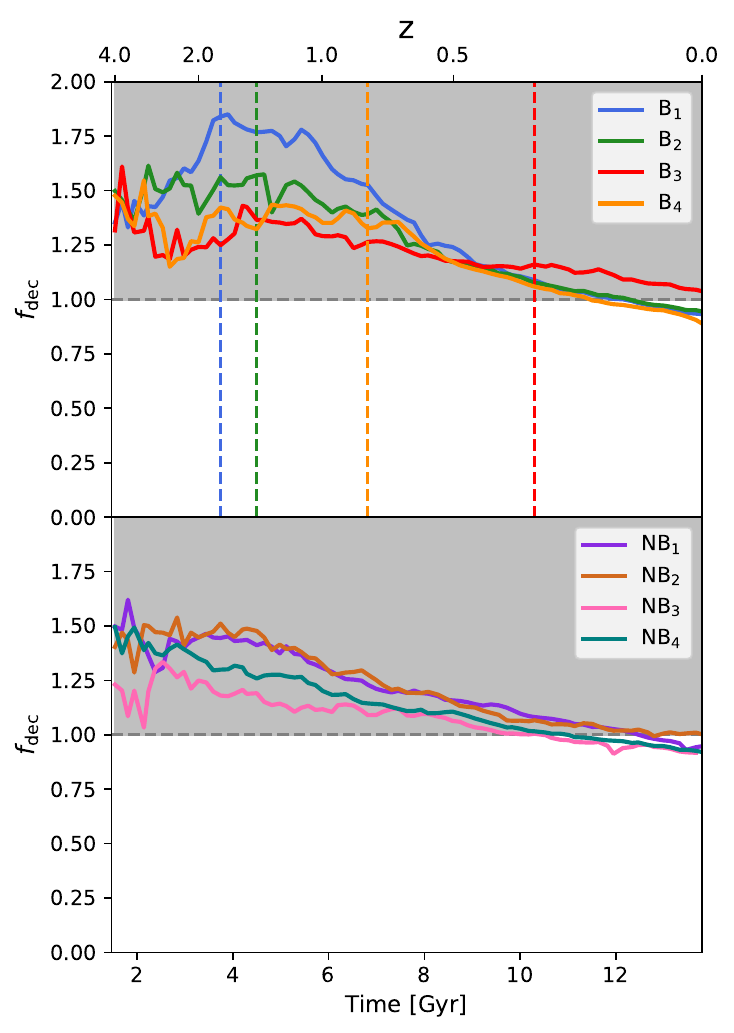}
    \caption{Time evolution of the parameter $f_{\rm{dec}}$, for the barred galaxies (\textit{upper panel}) and the unbarred galaxies (\textit{lower panel}). The vertical dashed lines in the upper panel are the time formation of each bar, $t_{\rm bar}$, and the shaded area denotes the instability region: $f_{\rm{dec}} > 1$.}
    \label{fig:fdec-evol}
\end{figure}

In order to investigate in more detail the parameters $f_{\mathrm{disc}}$ and $f_{\mathrm{dec}}$ and their ability to detect instabilities, we analyse their temporal evolution. 
Fig. \ref{fig:fdisc-evol} shows the evolution in time of  $f_{\mathrm{disc}}$, for the barred galaxies (upper panel) and the unbarred ones (lower panel). In the case of the barred galaxies, we observe that they enter the unstable region around $\sim 4\,{\rm Gyr}$. In the case of $B_1$ and $B_2$, this time approximately coincides with the formation time of the bar. However, 
$f_{\mathrm{disc}}$ falls below 1.1 between  $\sim 4$ and $6\,\mathrm{Gyr}$ before the bars of $B_3$ and $B_4$ are detected in the simulations. In all four cases, once the galaxy is unstable according to this criterion, it stays in this regime until $z=0$.  Consistently, the parameter remains stable, at all times, in our unbarred galaxies.

Fig. \ref{fig:fdec-evol} shows the time evolution of $f_{\mathrm{dec}}$, for the barred galaxies (upper panel) and the unbarred ones (lower panel). Unlike the case of $f_{\mathrm{disc}}$,
both the barred and unbarred galaxies remain in the unstable regime during  most of their evolution, and only around $z=0$ their $f_{\mathrm{dec}}$ values fall below the threshold value.
According to our findings, $f_{\mathrm{dec}}$ is not an adequate estimator for assessing the possibility of bar formation.
Predictions could improve if this is used in combination with $f_{\mathrm{disc}}$ \citep{Algorry_2017,Marioni_2022}.

\begin{figure}
    \includegraphics[width=1\columnwidth]{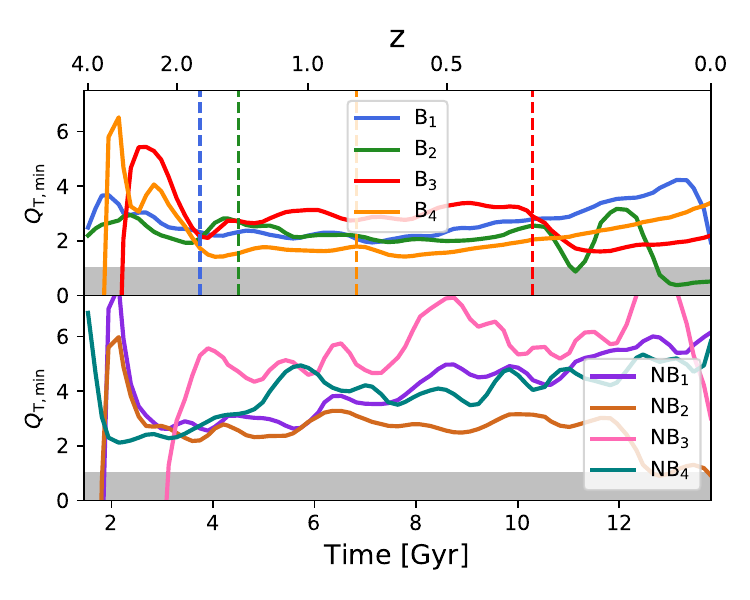}
    \caption{Evolution with time of the minimum value of the Toomre parameter, $Q_{\rm T, min}$ for the barred galaxies (upper panel), and the unbarred galaxies (lower panel). They shaded area represents the theoretical instability region ($Q_{\rm T,min}<1$).}
    \label{fig:Toomre_min}
\end{figure}

Another parameter  usually used in this type of studies is based on the Toomre local stability criterion, introduced by \cite{Toomre_1964},  which measures the stability of a rotating, self-gravitating disc. This criterion aids in predicting whether a disc will maintain its stability when exposed to the growth of axisymmetric disturbances, commonly manifesting as density waves. Such disturbances have the potential to trigger the development of non-axisymmetric structures like spirals or bars. The Toomre criterion considers the balance between rotational support, pressure forces, and self-gravity within the disc, quantified by the parameter: $Q_{\rm T} = (\sigma \kappa)/(3.36 G \Sigma)$, where $\sigma$ is the velocity dispersion, $\Sigma$ is the stellar surface density of the disc (estimated considering stars with a circularity parameter $\epsilon>0.7$), and $\kappa$ is the epicyclic frequency. The latter is estimated as:
\begin{equation}
\kappa^2(R)=\left(R\frac{\rm{d}\Omega^2}{\rm{d}R}+4\Omega^2\right)_{R},
\label{eq:epicycle}
\end{equation}
where $\Omega$ is the angular frequency (see Fig.~\ref{fig:phiRad_all_dehnen}).
According to theory, when $Q_{\rm T} > 1$, the stabilizing influence of rotation and pressure prevents the formation of a bar, ensuring the disc's stability by dissolving any axisymmetric perturbations.
On the other hand, if $Q_{\rm T} < 1$, the destabilizing effects of self-gravity come to the forefront, potentially amplifying axisymmetric perturbations and thereby facilitating the formation of structures such as spirals or bars.
Fig. \ref{fig:Toomre_min} displays the time evolution of the minimum Toomre parameter value, $Q_{\rm T,min}$ (found from the inspection or the radial profile of $Q_{\rm T}$ at different snapshots of the simulation) for barred galaxies (upper panel) and unbarred galaxies (lower panel). It can be observed that, although all galaxies present values greater than $Q_{\rm T,min} = 1$, for almost their whole evolution, the barred galaxies have lower values than the unbarred ones.
Thus, this parameter could be used to guide the tendency of galaxies to become unstable. 
It is worth noting that the failure to meet the theoretical threshold is anticipated, given the complex nature of the physical processes involved in bar formation  and not fully captured by theoretical estimations.
 
\subsection{External triggers: mergers, flybys and environment}\label{Sec:ExternalTriggers}

\begin{figure*}
    \centering
    \includegraphics[width=2\columnwidth]{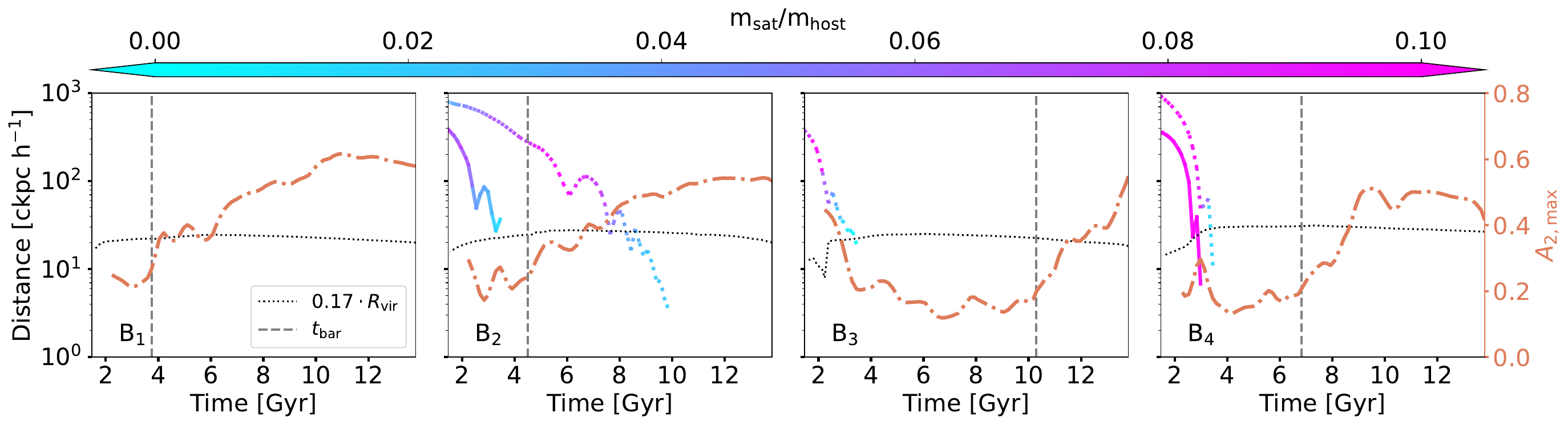}
    \includegraphics[width=2\columnwidth]{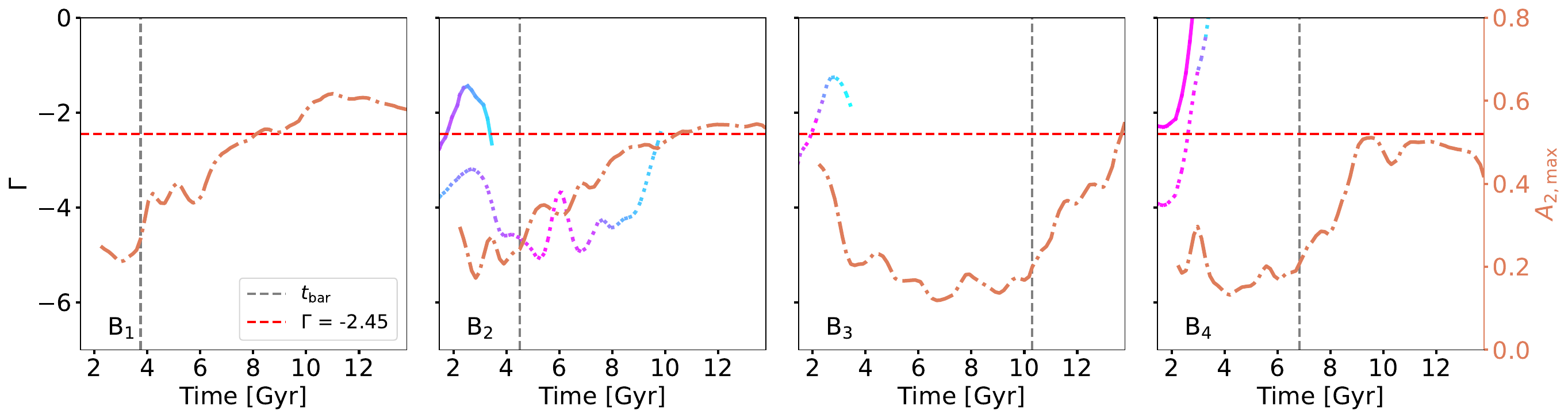}
    \caption{\textit{Upper panels}: Time evolution of the distance of various satellites to the galactic centre of their host, for the  barred galaxies. Each curve represents the distance evolution followed by a satellite, reaching its end when the merger event occurs; the thin dotted curve shows, for reference, the evolution of  $0.17 \times R_{\mathrm{vir}}$ as a function of time which corresponds, at $z=0$, to $\sim 30\, \mathrm{kpc}$. Each line is colour-coded according to the stellar mass ratio between the satellite galaxy and its host. \textit{Lower panels}: Scaled tidal index $\Gamma$, defined in the text, together with the threshold value of $\Gamma_{\rm t}=-2.45$ (horizontal dashed lines). The line styles for the satellites remain consistent with those used in the upper panels. In all panels, the dashed vertical line represents the bar formation time, $t_{\rm bar}$, for each barred galaxy. The dashed-dotted line depicts the time evolution of the bar strength, $A_{2, \rm max}$, with the corresponding values indicated in the right y-axis.}
    \label{fig:MERGERS}
\end{figure*}

The role of mergers and flybys in the formation and development of bars remains a subject of debate. From the analysis of Illustris, most barred galaxies at $z=0$ appear to have their bars triggered by external perturbations through mergers or flybys rather than forming through secular evolution \citep{Peschken_Lokas_2019}. Conversely, results from a cosmological zoom-in simulation that led to the formation of a Milky Way-like galaxy hosting a central bar in its central few kiloparsecs (ErisBH simulation, \citealt{Bonoli_2016}) demonstrate that the origin of bar formation is not necessarily linked to any specific tidal event; instead, it depends on the disc reaching a sufficient mass to sustain global non-axisymmetric modes, particularly for massive disc galaxies at $z\lesssim 1$. Additionally, a fly-by interaction could delay bar growth by increasing the stellar velocity dispersion, resulting in a more stable disc \citep{Zana_2018}. 
However, at high redshifts, tidal interactions with satellites tend to promote bar formation, which can subsequently drive the growth of a bulge or pseudo-bulge \citep{Guedes_2013}. While bulges are primarily triggered by major mergers at high redshifts \citep[e.g.][]{Fiacconi_2015}, pseudobulges grow gradually through non-axisymmetric secular disc instabilities \cite[e.g.][]{Devergne_2020}.
Furthermore, results obtained from the application of two convolutional neural networks to galaxies in the {\sc eagle} `reference' hydrodynamical simulation with stellar masses $\gtrsim 10^{10} \rm M_{\odot}$, spanning redshifts from $z = 1$ to $0$, indicate that barred galaxies undergo episodes of bar creation, destruction, and regeneration;  major mergers are connected to the destruction of bars, while minor mergers and accretion are associated with both the creation and dissolution of bars \citep{Cavanagh_2022}.

We have conducted a global analysis of the mergers experienced by each of the barred galaxies, through the use of the merger trees in the TNG simulations \citep{Rodriguez_Gomez_2015}, to provide an initial approach to this topic. The selection was based on the following criteria:
\begin{itemize}
    \item We only consider mergers that occurred at times later than $2.5$ Gyr, as at very early times the structure of the discs  is not yet well-defined (see also \citealt{Iza_2022}).
    \item We focus on mergers with maximum (stellar) mass ratios greater than $0.03$. Note that the stellar mass ratios (i.e. the ratio between the stellar mass of the merging galaxy and that of the host) vary with time. Consequently, we compute this ratio at each time step and identify cases where the maximum stellar mass ratio exceeds our predetermined threshold value instead of using its value at the time of the merger. In this way, we are certain that the mass ratio is not underestimated, which can occur as a result of the mass loss at times preceding the merger.
\end{itemize}
The results can be observed in the upper panels of Fig. \ref{fig:MERGERS}, where we show the distance between each selected satellite and the mass centre of the host, as a function of time, for barred galaxies. 
The colours represent the evolution of the mass ratios for the selected merger events. Higher ratios (more pinkish colours) are found at very early times (recall that we ignore times prior to $2.5\, \mathrm{Gyr}$) when the satellites are still quite far from the host. As satellites approach their host, they can lose stellar mass producing a decrease in the mass ratio, and also the host can gain stellar mass, producing the same effect. 
The satellites are followed until they are no longer identified as a distinct substructure in the merger tree. The formation time of each bar is indicated with a dashed grey vertical line.
Superimposed, we include the time evolution of $A_{2,\rm max}$ and present their corresponding values on the right y-axis. Additionally, for reference, we also show as a black dashed line the evolution of  $0.17\times R_\mathrm{vir}$, which corresponds to the central region of the galaxies (equivalent to $\sim 30 \, \mathrm{kpc}$ at $z=0$). We observe that our barred galaxies undergo a limited number of merger events: B$_1$ experiences no mergers based on our selection criteria; B$_2$ encounters two merger events, with one merging satellite dissolving close to the formation of the bar; B$_3$ has only one merger, an event that concludes significantly earlier than the bar's formation; and B$_4$ undergoes two mergers, with the satellites dissolving around $4\, \mathrm{Gyr}$ before the bar forms.

The information from the upper panels of this figure does not definitively indicate whether mergers act as triggers for bar formation: stellar mass ratios are quite small, and most mergers did not occur immediately before the formation of the bar, even though it may take several Gyr to form bars.

To better quantify the importance of these interactions, we calculate a scaled tidal index, as defined by \cite{Ansar_2023} in their equation (7), i.e.: 
\begin{equation}
    \Gamma = \log_{10} \left ( \frac{M_{\rm sat}/D^3_{\rm sat}}{V^ 2_{\rm circ,max}/GR^2_{\rm max}} \right )
\end{equation}
where $M_{\rm sat}$ is the mass of the satellite, $D_{\rm sat}$ is its distance to the host galaxy, $V_{\rm circ,max}$ is the maximum circular velocity, and $R_{\rm max}$ is its corresponding radius. 
They constructed this parameter based on the tidal index ($\Theta$) initially defined by \citet{Karachentsev-Makarov_1999}. The difference between both parameters is that $\Theta$ depends only on the mass of the satellite and its distance to the host, while $\Gamma$ introduces information about the dynamics of the host galaxy (the rotation curve). The authors make a choice for the value of $\Gamma$ from which an interaction can be considered significant: $\Gamma_{\rm t} = -2.45$ (see their paper for more details). The evolution over time of $\Gamma$ for each of the merged satellites is shown in the lower panels of Fig. \ref{fig:MERGERS}, keeping the same line style as in the upper panels.
We can see that the selected mergers might have had a tidal  influence on the host galaxy. Particularly, in the case of B$_2$, the tidal index of one of the satellites  becomes higher as the satellite approaches the host and exceeds the   threshold value $\Gamma_{\rm t}$ only $1\, \mathrm{Gyr}$ before the formation of the bar. A similar evolution of the tidal index is found for  B$_3$; however, the formation of the bar occurs significantly later than the merger, and it is therefore unlikely that it might have induced the formation of the bar. Finally, the tidal index of the two mergers experienced by B$_4$ increase rapidly reaching the threshold value around $3\, \mathrm{Gyr}$, suggesting that these might have affected the structure of the host, perhaps inducing an instability leading to bar formation. In this case, the bar forms at $\sim 7\, \mathrm{Gyr}$, and thus, if the mergers were the triggers of bar formation, it took about $5\, \mathrm{Gyr}$ for the system to form the bar. 
We also analysed the merging satellites of our simulated unbarred galaxies, and found that none of them experienced significant merger events throughout their evolution, and the $\Gamma$ parameter never exceeded $\Gamma_{\rm t}$. 

An analysis of flybys has also been carried out, focusing specifically on those that might have had an impact on bar formation: for barred galaxies, the analysis focuses on times before $t_{\rm bar}$, while for unbarred galaxies, we consider all times until $z=0$. Flybys were identified as satellites that approach, at any time during their evolution, to at least $50 \, \rm ckpc$ from the host (excluding those that have merged, which enter in our merger list). We have used a minimum stellar mass ratio of $0.03$, similarly to our merger analysis. Based on these criteria, only B$_4$ experienced a flyby and, for the unbarred galaxies, NB$_2$ (2 flybys) and NB$_3$ (1 flyby) have had such events. None of the flybys has a tidal index above our threshold value to be considered significant, except for one of the flybys of NB$_2$ with $\Gamma>-2.45$ right before the time where $A_2$ starts to grow, at $\sim 12.5\, \mathrm{Gyr}$. This, together with other indications discussed throughout the text, suggests that NB$_2$ might be initiating the process of developing a bar, and this might be related with the close passage of a satellite galaxy. 

\begin{figure}
    \centering
    \includegraphics[width=1\columnwidth]{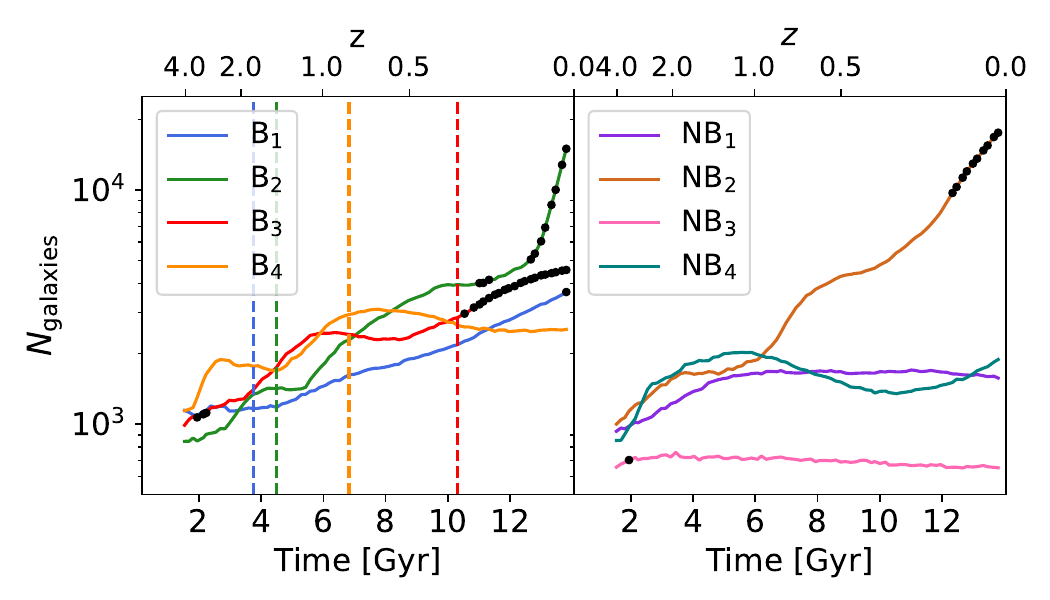}
    \caption{Time evolution of the density of the environment for our barred and unbarred samples, measured as the number of neighbouring galaxies located within a sphere of $1\,{\rm cMpc}$ radius around the host.}
    \label{fig:Environment}
\end{figure}

Finally, we examine the possibility that environment might affect bar formation. In Fig.~\ref{fig:Environment} we show the environmental density of barred and unbarred galaxies as a function of time. Environmental density has been estimated as the number of galaxies within $1$ cMpc radius around the host. 
Black dots indicate moments when galaxies are recorded as satellites of their host haloes (similar to Fig. \ref{fig:a2t}).
We find that the barred galaxies live in denser environments compared to the unbarred sample, except for NB$_2$ which exhibits a fast growth in the number of galaxies within the considered radius, and interestingly, this is the galaxy that seems to be starting to form a bar recently, at $\sim 12.5\, \mathrm{Gyr}$ (see Fig.~\ref{fig:a2t}).
Despite differences in the environmental densities, both the barred and unbarred galaxies are central (and not satellites of a larger system) during most of their evolution. However, for all barred galaxies the environmental densities increase with time, but remain relatively constant for the unbarred sample (except for NB$_2$, as discussed above). This means that barred galaxies might be suffering environmental changes, and it is still to be seen whether this can have an impact in the triggering of  instabilities and bar formation. 

\section{Discussion} \label{Sec:Discussion}

In this Section we discuss our results and compare them to those presented in other studies based on numerical simulations. 
Comparisons primarily focus on simulations within a cosmological context, with additional consideration given to results from models of isolated halo-disc systems, if appropriate.
When available, we also investigate the level of agreement between our findings for bar properties and observational estimates.

In general terms, our results are consistent with those based on  cosmological simulations. First, we find that barred galaxies exhibit a faster growth at early times compared to their unbarred counterparts, and posses more massive stellar components, which are achieved as the bar develops even prior to the bar formation time ($t_{\rm bar}$).
These results are consistent with the findings  of \cite{Algorry_2017, 
RosasGuevara_2022} and \cite{Izquierdo-Villalba_2022}, based on the {\sc eagle} and {\sc tng} simulations. The bar properties obtained in our work -- bar length, bar strength, corotation radius and parameter $\mathcal{R}$ --  are also compatible with these works.
Our result that unbarred galaxies have low stellar to dark matter mass ratios is also consistent with results from isolated galaxy simulations \citep{Athanassoula_2013, Reddish_2022, Bland-Hawthorn_2023}.

Even though we have a small sample of four barred galaxies, they exhibit a wide range in formation times, from $\sim 3.74$ to $\sim 10.3 \, \mathrm{Gyr}$. This is consistent with the bar formation times obtained in  \cite{Izquierdo-Villalba_2022}
and in \cite{Algorry_2017}, although in this last case they found that very few of their galaxies have measurable bars that formed in the first $5\, \mathrm{Gyr}$. Our four bars are strong, and have a  relatively quick growth since they are first identified at $t_\mathrm{bar}$, similarly to the findings of \cite{Algorry_2017} for their strong bar sample.
It is interesting to note that, from our four barred galaxies,  the two galaxies with lower $t_\mathrm{bar}$ (i.e the two that formed their bars earlier) grow over longer timescales compared to the two galaxies whose bars developed later ($z<1$).

Irrespective of their formation time, these bars exhibit low gas content in the inner regions at $z=0$, except for B$_3$ as previously discussed, due to substantial consumption by high levels of star formation. This aligns with the findings of \cite{Algorry_2017}, indicating that bars tend to develop more frequently in systems that are presently gas-poor. It is crucial to consider the specific moment when gas content is assessed when presenting such correlations. Notably, our barred galaxies initially have higher gas fractions, leading to conflicting trends compared to isolated galaxy  simulations. These trends typically suggest a reduction in bar strength with larger gas fractions as the central mass concentration becomes more massive and induces vertical stellar motion \citep{Athanassoula_2013, Seo_2019}. In contrast, our barred galaxies exhibit a pronounced and sustained growth of the CMC over time, resulting from a combination of high initial gas levels and subsequent increased rates of star formation.

The evolution of the bar strength ($A_2^{\rm max}$) of our barred galaxies (upper panel Fig.~\ref{fig:a2t}), exhibits a pattern akin to the bar growth observed in the non-cosmological simulations conducted by \cite{KatariaShen_2022} and differs from those obtained in the other disc-halo models described in Section~\ref{Sec:EvolBarLengthPatternSpeed}. From our findings, we can conclude that the internal angular momentum transfer is key in determining the bar growth and that dynamical friction exerted by the halo also plays an important role \citep{ShlosmanNoguchi_1993, Athanassoula_2014}.
A deeper analysis of the angular momentum transfer and the relation between bar formation and halo properties in the context of cosmological simulations (L\'opez et al., in preparation) will be carried out in follow-up works. 

Our analysis of the instability parameters showed that no single parameter can predict the formation of a bar. In particular, $f_\mathrm{disc}$ is the one where a correlation between the time where the system becomes unstable ($f_\mathrm{disc}<1.1$) and the identification of the bar is more clear. 
On the other hand, $f_\mathrm{dec}$  sets our galaxies -- both barred and unbarred -- into the unstable regime ($f_\mathrm{dec}>1$) for all times, with the exception of the unbarred galaxies very near $z=0$. When both instability conditions ($f_\mathrm{disc}<1.1$ and $f_\mathrm{dec}>1$) are considered together at $t_{\rm bar}$ (for barred galaxies) and at $z=0$ (for unbarred galaxies), our sample is properly divided into systems that do/do not posses a bar, consistent with the results of  \cite{Algorry_2017} (note, however, that at previous times this does not hold). 
Our findings suggest that both the local and global stability of a disc-halo system plays a crucial role in determining the likelihood of bar formation. However, the precise threshold for determining (in)stability may not fully capture the complexity inherent in galaxies formed within a cosmological context.
Finally, the Toomre instability criterion predicts systematically different values for the barred and unbarred galaxies, although the threshold value commonly assumed to separate into stable and unstable systems does not work in our sample.
This is in line with other works that also found that the Toomre parameter alone can not predict the formation of a bar, and generally this is used in combination with, e.g., the swing amplification parameter \citep[e.g.][]{Spinoso_2017}.

Among the galaxies that formed  bars early on, B$_2$ is found to have possibly formed due to a merger that occurred $1$ Gyr before $t_{\rm bar}$, and B$_1$ shows no signs of mergers under the chosen criteria. This contrasts with the results of \cite{Peschken_Lokas_2019} who find that, at high redshift, bars are more likely to form due to interactions and later tend to disappear through secular evolution, meanwhile the bars in our sample are able to survive until $z=0$. 
The other two barred galaxies experienced mergers with high tidal index, but they merged $>4\, \mathrm{Gyr}$ prior to $t_\mathrm{bar}$, which suggests that mergers were not linked to the formation of the bars.
We find no clear correlation between the environment where galaxies live and the presence or absence of a bar, although all unbarred galaxies (with the exception of NB$_2$) were found to be in less dense environments compared to the barred galaxies. However, at times similar of the formation of the bars, the environmental densities are still moderate and similar to those of the unbarred systems. 
This aligns with the conclusions drawn in the study done by \cite{Sarkar_2020}, who utilised SDSS data and found no notable impact of the environment on bar formation.

While comparisons between observational and simulation studies should always be interpreted with caution due to the very different techniques used to derive the bar quantities (which might be influenced by observational biases or numerical aspects), comparisons are useful and can provide clues on the formation processes of bars. In our work, we find that the simulated bars are in general in good agreement with observational results. 
In particular, bar lengths are well within in the range of observed bars, as well as $\mathcal{R}$ values (e.g. \citealt{Gadotti_2011, Aguerri_2015, Geron_2021}, based on the SDSS, CALIFA and MaNGA surveys).
The recent observational study  of \citet{Geron_2023}, who analyse 225 barred galaxies using IFU data from MaNGA, also reported measurements of the bar pattern speed for their sample. They obtained a median value of $\Omega_{\rm bar}= 23.36^{+9.25}_{-8.1} \,{\rm km} \,{\rm s}^{-1}\,{\rm kpc}^{-1}$ for the whole sample, which comprises systems with a broad range of stellar masses (log($M_*/$M$_\odot$) between $\sim 10$ and $11.5$). For stellar masses similar to those of our simulations (log($M_*/$M$_\odot) \sim 10.7$), the corresponding median is about $25 \,{\rm km} \,{\rm s}^{-1}\,{\rm kpc}^{-1}$, which is somewhat lower than but still consistent with the pattern speeds obtained in the simulations. 

Finally, our findings related to the barred galaxies exhibiting stronger star formation episodes at early times compared to the unbarred sample (similar to the result obtained by \citealt{RosasGuevara_2022}, for galaxies in the TNG100 simulation) seem consistent with the observational study of \cite{FraserMcKelvie_2020}, who  also found that the star formation histories of barred galaxies peak at earlier times than those of unbarred systems. 

\section{Summary and conclusions}\label{sec:conclusions}

We used the magnetohydrodynamical, cosmological simulations of the IllustrisTNG project, in particular TNG50, to study the formation of galactic bars. For this purpose, we selected eight galaxies, four of which have bars at $z=0$, based on the classification of barred and unbarred galaxies of \cite{RosasGuevara_2022}. The simulated galaxies have present-day virial masses in the range  $3.2 \times 10^{11} {\mathrm{M}}_{\odot}$ to  $1.5 \times 10^{12} {\mathrm{M}}_{\odot}$. 
We comparatively studied the evolution of the barred and unbarred galaxies, with the aim of identifying the physical processes that can trigger instabilities and induce the formation of bars.

We used a Fourier decomposition of the stellar distributions to identify bars and calculated their bar length ($R_{\mathrm {bar}}$), strength ($A_2^{\mathrm{max}}$) and pattern speed ($\Omega_{\mathrm{bar}}$). We found, at $z=0$, the following  values for these quantities: $R_\mathrm{bar}$ in the range  $\sim 3-5~\mathrm{kpc}$,  $A_2^{\mathrm{max}}$ between $0.41$ and $0.58$ -- setting them into the strong bar category--  and $\Omega_{\mathrm{bar}}\sim 30~\mathrm{km\, s^{-1}\, kpc^{-1}}$. We also calculated the corotation radii ($R_{\rm{CR}}$), which are in the range $\sim 5-6~\mathrm{kpc}$, and the parameter $\mathcal{R}=R_{\rm{CR}}/\rm{R}_{\rm{bar}}$  which allows to classify bars into slow or fast rotators. From the four simulated bars, half of them are, at $z=0$, slow bars and the other half are fast.

We analysed the time evolution of the bars, and calculated their  formation times ($t_\mathrm{bar}$). We found a great variety of formation times in our four simulated galaxies, between $3.74~\mathrm{Gyr}$ and $10.3~\mathrm{Gyr}$. 
All simulated bars grow steadily with time since their formation, both in terms of their $A_2^\mathrm{max}$ parameters and lengths,  and even those formed very early on survive until the present time. The evolution of the pattern speed is more complex, with a general behaviour of an increase right after the formation of the bars and a subsequent decrease until $z=0$.  

The comparative analysis of the evolution of the galaxies with and without bars allowed to identify some differences which we summarize below:
\begin{itemize}

\item The circular velocity curves of the barred galaxies are, in the central regions, more peaked compared to those of the unbarred systems. This is due to the contribution of stars in the centre of the galaxies, which largely dominate over the dark matter component. The unbarred galaxies have circular velocity curves that are smoother in the central regions, and the stellar mass relative to  dark matter mass is lower compared to the case of barred galaxies. This behaviour is found at all times and, in the case of galaxies with bars, is present even before the bars form. This suggests that the characteristics of the halo and the relative contribution of stars and dark matter in the inner regions play  a role in bar formation, as found by previous studies. 

\item When examining the spin of the halo within the inner $10$ ckpc, we observed a consistent pattern where bars originate due to angular momentum transfer from the disc to the halo (only the behaviour of B$_3$ differs from the others). 
Nevertheless, even though the initial value of $\lambda_{10 \rm \, ckpc}$ is quite similar for all the galaxies and aligns with the low spin values for bar formation, as suggested by previous studies, certain galaxies do not undergo bar formation, and the total spin of their halo remains relatively constant over time. This suggests, on the one hand, that the initial spin value alone is not sufficient to determine whether a bar forms or not. On the other hand, it underscores the importance of the radial distribution of spin for distinguishing between barred and unbarred galaxies.

\item The differences detected in the circular velocity curves between barred and unbarred galaxies are also noticeable from the ratio
$f_\mathrm{star}^{1\mathrm{ckpc}}\equiv M_\mathrm{star}/(M_\mathrm{star}+M_\mathrm{DM})$ with $M_\mathrm{star}$ and $M_\mathrm{DM}$ measuring the stellar and dark matter masses within the inner $1~\mathrm{ckpc}$.
The $f_\mathrm{star}^{1\mathrm{ckpc}}$ reveal that barred galaxies exhibit a higher fraction compared to unbarred ones, despite having similar initial values. When analysing the gas fraction relative to dark matter$f_\mathrm{gas}^{1\mathrm{ckpc}}$, it was observed that barred galaxies harbour larger gas reservoirs in the inner regions in contrast to unbarred ones. This behaviour aligns with the high rates of star formation found in barred galaxies, contributing to the increased stellar content in the inner regions. On the other hand, unbarred galaxies show a more gradual history of star formation, with the exception of NB$_2$, which exhibits more pronounced peaks.

\item The CMC parameter, estimated from our simulations as the sum of stars and gas within the inner $1\,{\rm ckpc}$ confirms the aforementioned trends based on the relative contribution of baryons with respect to dark matter in the inner regions and the shape of the velocity curves. Although the CMC increases monotonically with time for both barred and unbarred galaxies, the former experience a higher rate of increase than the latter, consistent with the fact that they have initially more gas content. This results in larger CMC values at $z=0$ for barred galaxies, and contradicts earlier findings drawn from simulations of isolated disc-halo systems, which proposed that a compact, centrally concentrated component would provide stability against bar formation. Nevertheless, the cosmological simulations employed in this study, along with those presented by other researchers, underscore the highly intricate nature of the bar formation process, which involves gas inflows and outflows affecting the radial distribution of the baryonic component.

\item From our merger/flyby analysis, we conclude that while these events might trigger the formation of bars, similar interactions occur in galaxies that form/do not form a bar. It is therefore extremely difficult to  disentangle the effects of mergers and interactions on the stability properties of galaxies formed within the standard cosmological model and their ability to form bars, and other properties probably play a role as well. We also performed an analysis of environment and found that barred galaxies inhabit denser environments, and present stronger environmental evolution compared to the unbarred ones, which (with the exception of one galaxy) stay as relatively isolated systems with low environmental densities. This might suggest a connection between the large-scale structure around galaxies and the development of instabilities, presumably in connection with a  richer merger/interaction history. 

\item We investigated various stability parameters used in the literature, in order to test whether any of them can predict the formation of bars. In particular, we used estimations from \cite{Algorry_2017} ($f_\mathrm{disc}$ and $f_\mathrm{dec}$; the former being a proxy of the instability parameter introduced by \citealt{Efstathiou_1982}), and \cite{Toomre_1964} ($Q_{\rm T}$; in particular the minimun of the radial profile for each time). 
According to our results, only the $f_\mathrm{disc}$ parameter is able to separate galaxies into stable (the unbarred galaxies) and unstable (the barred galaxies). However, the times at which the barred galaxies become unstable do not always relate to the times where bars are detected in the simulations. A similar situation is observed in the case of the Toomre parameter, and the standard threshold to separate galaxies according to their stability properties does not work for our sample, although barred galaxies do have systematically lower $Q_{\rm T}$ values than the unbarred ones. 
\end{itemize}

The cosmological simulations used here, as well as those presented by other authors, highlight that the process of bar formation is extremely complex, and is not captured when a single parameter is considered to evaluate the disc instability.
While the phenomenological connection between the evolution of different properties we have presented provides insights into the bar formation process, it does not offer a clear explanation for the physical cause or particular combination of parameters that trigger the disc instability. A comprehensive examination of the radial distribution of the central galaxy's components, coupled with a detailed analysis of gas inflows and outflows, is imperative to draw definitive conclusions about the influence of various parameters as indicators of bar formation.

\section*{Acknowledgements}

We thank Francesca Fragkoudi and Dimitri Gadotti for fruitful discussions and comments. We are grateful to the anonymous referee for useful suggestions and comments that contributed to enhancing the paper. SC and SAC acknowledge funding from {\it Consejo Nacional de Investigaciones Científicas y Tecnológicas}  (CONICET, PIP-2876 and PIP KE3-11220210100595CO), and {\it Agencia Nacional de Promoci\'on de la Investigaci\'on, el Desarrollo Tecnol\'ogico y la Innovaci\'on} (Agencia I+D+i, PICT-2018-3743 and PICT-2021-GRF-TI-00290). SAC acknowledges support from the {\it Universidad Nacional de La Plata} (G11-150), Argentina. PDL has received financial support from the European Union's HORIZON-MSCA-2021-SE-01 Research and Innovation programme under the Marie Sklodowska-Curie grant agreement number 101086388 - Project acronym: LACEGAL.

\section*{Data Availability}

The TNG simulations, including TNG50, are publicly available at https://www.tng-project.org/.  
The scripts and plots used in this article will be shared on reasonable request to the corresponding author.


\renewcommand{\bibname}{REFERENCES}
\bibliographystyle{mnras}
\bibliography{references} 


\bsp	
\label{lastpage}
\end{document}